\documentclass[12pt,preprint]{aastex}
\usepackage{amsmath}
\usepackage{amsfonts}
\usepackage{dsfont}
\usepackage{amsxtra}
\usepackage{hyperref}
\usepackage{amssymb}

\usepackage{graphicx,epsfig}
\usepackage{dcolumn}% Align table columns on decimal point
\usepackage{bm}% bold math

\newcommand{\picsizetwo}{0.4}
\newcommand{\picsize}{0.5}

\newcommand{\be}{\begin{equation}}
\newcommand{\ee}{\end{equation}}
\newcommand{\bea}{\begin{eqnarray}}
\newcommand{\eea}{\end{eqnarray}}
\newcommand{\beaa}{\begin{eqnarray*}}
\newcommand{\eeaa}{\end{eqnarray*}}
\newcommand{\ba}{\begin{array}}
\newcommand{\ea}{\end{array}}
\newcommand{\bi}{\begin{itemize}}
\newcommand{\ei}{\end{itemize}}
\newcommand{\ben}{\begin{enumerate}}
\newcommand{\een}{\end{enumerate}}

\newcommand{\bra}{\langle}
\newcommand{\ket}{\rangle}
\newcommand{\ra}{\rightarrow}
\newcommand{\lra}{\longrightarrow}

\newcommand{\td}{\tilde}

\newcommand{\lb}{\label}

\newcommand{\al}{\alpha}

\newcommand{\dl}{\delta}
\newcommand{\Dl}{\Delta}

\newcommand{\Om}{\Omega}
\newcommand{\om}{\omega}
\newcommand{\sm}{\sigma}

\newcommand{\Fermi}{\textsl{Fermi} }

\begin{document}

\title{ Statistics of gamma-ray point sources \\
below the \textsl{Fermi}  detection limit 
\bigskip
\bigskip
}

\author{Dmitry Malyshev\altaffilmark{1} and David W. Hogg}
%\email{dm137 at nyu.edu}

%\author{Ilias Cholis}
% \email{ijc219@nyu.edu}

%\author{Joseph Gelfand}
 %\email{jg168@astro.physics.nyu.edu}

 \affil{
 \bigskip
 Center for Cosmology and Particle Physics, New York University \\
 4 Washington Place, Meyer Hall of Physics, New York, NY 10003, USA}

\altaffiltext{1}{dm137 at nyu.edu\\
On leave of absence from ITEP, Moscow, Russia, B. Cheremushkinskaya 25}

%\date{\today}

\begin{abstract}
%\newpage

An analytic relation between the statistics of photons in pixels
and the number counts of multi-photon point sources is 
used to constrain the distribution of gamma-ray point
sources below the \textsl{Fermi} detection limit at energies above 1 GeV
and at latitudes below and above $30^\circ$.
The derived source-count distribution is consistent with the distribution
found by the \textsl{Fermi}  collaboration based on the first \textsl{Fermi} point source catalogue.
In particular, we find that the contribution of resolved and unresolved active galactic nuclei (AGN)
to the total gamma-ray flux is below 20 to 25\%.
In the best fit model, the AGN-like point source fraction is $17\pm 2$\%.
Using the fact that the Galactic emission varies across the sky
while the extra-galactic diffuse emission is isotropic,
we put a lower limit of 51\% on Galactic diffuse emission
and an upper limit of 32\% on the contribution from extra-galactic weak sources,
such as star-forming galaxies.
Possible systematic uncertainties are discussed.

\end{abstract}


%\pacs{
%98.70.Rz %gamma-ray sources; gamma-ray bursts 
%98.54.Cm %Active and peculiar galaxies and related systems (including BL Lacertae objects, blazars, Seyfert galaxies, Markarian galaxies, and active galactic nuclei) 
%95.85.Pw %gamma-ray 
%02.70.Rr %General statistical methods 
%}

%\maketitle
%\end{titlepage}

%\newpage

%\tableofcontents

% begin body

\section{Introduction}

Faint gamma-ray point sources cannot be detected individually,
however their presence affects the statistics of photons across the sky.
One can use the observed statistics of photons to infer some general properties
about the population of point sources below the \textsl{Fermi} detection limit.
The idea of using the statistics of photon counts or intensity maps in the study
of faint point sources is familiar in radio %, IR, 
and X-ray observations
\citep[e.g.,][]{1957PCPS...53..764S, 1974ApJ...188..279C, 
1974MNRAS.166..329S, 1993A&A...275....1H, 2002ApJ...564L...5M, 2011arXiv1101.0256S}.
A closely related subject is the use of extreme statistics to constrain the non-Gaussianity
of cosmic microwave background data, e.g., \cite{2011arXiv1102.5707C} and references therein.
In this paper, we extend the prior analysis 
\citep{2010MNRAS.405.1777S} and
use the points in cells statistics
to understand the population of gamma-ray point sources.

In a larger context, the problem is to separate different sources of gamma-ray emission.
The standard strategy is either to use templates \citep{2010ApJ...717..825D} 
that trace the sources or else to simulate the cosmic ray propagation and gamma-ray production using, 
e.g., Galprop \citep{2007ARNPS..57..285S,2009arXiv0907.0559S}.
In this paper we would like to adopt a different methodology: 
we assume some general properties of the sources
in order to obtain model-independent constraints on the contribution from these sources.

We separate the sources based on their statistical properties.
In the paper we consider three sources of gamma-rays at high latitudes.
The first source is a diffuse source that varies over large angles.
The main contribution to this source comes from the Galactic diffuse emission
($\pi^0$ production, ICS photons, and bremsstrahlung) 
and, possibly, a population of faint Galactic point sources, such as millisecond pulsars
\citep{2010JCAP...01..005F, 2010ApJ...722.1939M, 2010arXiv1011.5501S}.
We will call this source non-isotropic Galactic diffuse emission.
It puts a lower limit on the actual Galactic emission.

The second source corresponds to an isotropic distribution of gamma-rays,
i.e., the statistics of photons across the sky is consistent with the Poisson distribution for this source.
The isotropic flux has contributions from the homogeneous part of the Galactic diffuse emission,
from diffuse intergalactic emission \citep[references can be found in][]{2010arXiv1012.3678S},
and from a population of very weak extragalactic sources
that on average emit much less than one photon during the time of observation,
such as star-forming galaxies \citep[e.g.,][]{2002ApJ...575L...5P, 2010ApJ...722L.199F},
starburst galaxies \citep{2007ApJ...654..219T}, galaxy clusters \citep{2003ApJ...594..709B}, etc.
The value of the isotropic emission puts an upper limit on the gamma-ray flux from very weak extra-galactic sources,
as well as an upper limit on the contamination from cosmic rays \citep{2010PhRvL.104j1101A}.

The third source is a population of point sources
modeled by a broken power-law source-count distribution.
We assume that the point sources are distributed homogeneously over the sky.
The statistics of photons coming from these point sources
has a non-trivial form (derived in Appendix \ref{sect:stat})
different from the Poisson statistics.
These sources model a population of AGN-like point sources
\citep{1993ApJ...410L..71S,1993MNRAS.260L..21P, 1995ApJ...452..156C,
1996ApJ...464..600S,
2000MNRAS.312..177M, 2010arXiv1012.1247A, 2011arXiv1103.3484N}.

We note that in our approach it is impossible to separate the isotropic part of 
Galactic emission from the diffuse extragalactic background (EGB).
Consequently, all results will be quoted either with respect to the total gamma-ray flux
or in absolute values.
We find that at high latitudes (below and above $30^\circ$) for energies above 1 GeV
the contribution of AGN-like point sources (both resolved and unresolved) 
to the total gamma-ray flux is $17\% \pm 2$\%,
the contribution from Galactic diffuse emission is above 51\% and the contribution from
extra-galactic weak sources is below 32\%.
Using these values, we estimate that the contribution of unresolved AGN-like point 
sources to the EGB model in \cite{2010PhRvL.104j1101A}
is below $\sim 25\%$, which is consistent with \cite{2010ApJ...720..435A}.

The paper is organized as follows.
In Section \ref{sect:model}, we describe a general model and a fitting algorithm.
In Section \ref{sect:data}, we present an analysis of the \Fermi data.
Section \ref{sect:discus} has discussion.
In Appendix \ref{sect:stat}, we derive the statistics of photons coming from a population of point sources.
In Appendix \ref{sect:LS}, we determine a model for non-isotropic Galactic diffuse emission.
In Appendix \ref{sect:PIX}, we repeat the data analysis of Section \ref{sect:data}
for different pixel sizes.

\section{Model}
\lb{sect:model}

In this section we describe the model that we later use to fit the \textsl{Fermi} gamma-ray data.
At first, we present an analytic relation between the counts of photons in pixels
and the statistics of point sources.
Then we take into account the detector point spread function (PSF) and 
describe a model of Galactic diffuse gamma-ray emission.
The model that we present in this section is rather general.
An application of this analysis for the \Fermi data is presented in Section \ref{sect:data}.

\subsection{Statistics of Photon Counts and Point Sources}

We will assume a pixelation of the sphere with pixels of equal area.
Usually a model of gamma-ray emission is represented in terms of 
the expected number $x_p$ of gamma-rays in pixel $p$.
The problem is that the number and the positions of weak point 
sources are not known.
The position of a point source can be found only by a ``detection''
of the source.
However, if one is interested only in the total number of 
sources of a given flux, then a ``detection'' of sources is not necessary.
In this paper, we use the statistics of photon counts in pixels,
in order to find the source-count distribution 
of point sources without actual identification of the sources.

In order to find the statistics of photon counts in pixels we calculate
the total number of pixels $n_k$ that contain $k$ photons.
In this calculation, the information about the position of the pixels on the sky
is not preserved, but the remaining information may be sufficient
to constrain general properties of the population of sources.
There are two competing conditions for this method to work.
On the one hand, there should be sufficiently many pixels to determine
the statistics of photon counts,
while, on the other hand, the size of the pixels cannot be too small compared to
the PSF, otherwise the statistics corresponding to the presence of 
point sources cannot be distinguished from non-isotropic diffuse emission.

If the total number of pixels is $N_{\rm pix}$
and $n_k$ is the observed number of pixels with $k$ photons,
then we can estimate the probability to find 
$k$ photons in a pixel as 
\be
\lb{eq:pdf_obs}
p_k = \frac{n_k}{N_{\rm pix}}.
\ee
For large $N_{\rm pix}$, the statistical uncertainty of $n_k$ is
approximately $\sqrt{n_k}$ (there is a small correction due to the fact
that the total number of pixels is fixed, i.e., the process is multinomial).
If there are no point sources and all the emission comes from an isotropic
diffuse source, then the probability distribution $p_k$ is
the Poisson probability of getting $k$ photons given the mean rate.
In the presence of point sources the probability distribution is more complicated
than the Poisson distribution.
In the context of X-ray point sources,
the probability distribution is called a $P(D)$ distribution 
(or a $P(D)$ diagram)
and is usually computed with the help of Monte Carlo simulations
\citep[e.g.,][]{2002ApJ...564L...5M}.

In this paper we find an analytic relation between the probability distribution
of photon counts in pixels and the source-count distribution.
We will use this relation to find a source-count distribution
that has the best fit to the observed probability distribution 
determined in Equation (\ref{eq:pdf_obs}).

In calculations we use the method of probability generating functions
\citep[an introduction can be found in, e.g.,][Section 3.6]{Hoel1971}.
For a given discrete probability distribution $p_k,\;k=0,1,2\ldots$,
the corresponding generating function is defined as 
a power series in an auxiliary variable $t$
\be
\lb{eq:pgf_def}
P(t) = \sum_{k = 0}^{\infty} p_k t^k.
\ee
If the generating function $P(t)$ is known,
then the probabilities $p_k$ can be found by picking the coefficient
in front of $t^k$, or, equivalently, by differentiating with respect to $t$
\be
p_k = \left. \frac{1}{k!} \frac{d^k P(t)}{dt^k} \right|_{t = 0}.
\ee
An important property of probability generating functions
is that a sum of two independent random variables corresponds
to a product of the corresponding probability generating functions
\citep{Hoel1971}.
For example, if there are two independent sources of gamma-rays
on the sky, e.g., Galactic and extragalactic,
then the probability generating function 
for the photon counts with the two sources
overlaid is given by the product of probability generating functions
corresponding to the two sources
(see Appendix \ref{sect:stat} for more details).

%One can also use circles at random positions instead of pixels.
%The problem is that for large number of circles many of the circles will overlap.
%Since the photon counts in overlapping circles are correlated,
%the statistical uncertainty is more complicated than a simple estimate $\sqrt{n_k}$.

Let $x_m$ denote the average number of sources inside a pixel that emit exactly
$m$ photons during the time of observation.
Using the sum-product property of probability generating functions
\citep[][Section 3.6]{Hoel1971},
we derive in Appendix \ref{sect:stat} the following relation between the 
probability generating function for the photon counts
and the expected number $x_m$ of $m$-photon sources 
\be
\lb{eq:genf}
\sum_{k=0}^\infty p_k t^k = \exp\left({\sum_{m = 1}^\infty (x_m t^m - x_m)}\right).
\ee
This formula provides an analytic relation between the
expected numbers of $m$-photon sources
and the statistics of photons in pixels.
The probabilities $p_k$ to observe $k$ photons in a pixel are determined 
by expanding the right hand side of this equation and picking 
the coefficient in front of $t^k$.

If we substitute $t = e^{i \om}$,
then Equation (\ref{eq:pgf_def}) becomes a 
discrete Fourier transform analog of the probability 
characteristic functions \citep[e.g.,][]{Hoel1971}.
The probability characteristic functions were extensively used
in the study of radio point sources below detection limit 
\citep[e.g.,][]{1957PCPS...53..764S}.
Similarly to probability generating functions,
the characteristic function for a sum of two independent random
variables corresponds to a product of the two characteristic functions.

The statistics of point sources can be described in terms of the differential source counts
as a function of flux $S$,
denoted by $dN / dS$.
One of the simplest forms for the source-count distribution is
a broken power-law
\be
\lb{eq:dNdS_power}
\frac{d N}{d S} \sim 
\left\{
\ba{l}
S^{-n_1}, \qquad S > S_{\rm break} \\
S^{-n_2}, \qquad S < S_{\rm break},
\ea
\right.
\ee
where we have to require $n_1 < 2$ and $n_2 > 2$
in order to have a finite total flux%
\footnote{
A benchmark model is provided by a static universe filled homogeneously with sources that have intrinsic luminosity $L_0$.
The number of sources with flux larger than $S = \frac{L_0}{4\pi R^2}$ is $N(S' > S) \sim R^3 \sim S^{-1.5}$.
The differential number count is $dN / dS \sim - S^{-2.5}$.
Index $n = 2.5$ provides a benchmark value for differential source count at large $S$.
%At small fluxes the distribution should have an index $n < 2$, otherwise the total flux is divergent for $S \ra 0$.
A break or a cutoff at small flux (large distance to the source) is necessary
given the finite size of the visible universe.
}.

Let $S$ be an average flux from a source 
(normalized to give the number of photons during the time of observation).
Then the probability to observe $m$ photons from this source is
\be
p_m(S) = \frac{S^m}{m!} e^{-S}.
\ee
If the sources are distributed isotropically, 
then the average number of $m$-photon sources inside a pixel is
given by the Poisson probabilities for sources with flux $S$ to emit $m$ photons
\be
\lb{eq:xm_noPSF}
x_m = \frac{\Om_{\rm pix}}{4\pi} \int_0^\infty dS \frac{dN}{dS}(S) \frac{S^m}{m!} e^{-S},
\ee
%where $S$ is normalized to the total flux during the time of observation and 
where $\Om_{\rm pix}$ is the angular area of the pixel.

\subsection{PSF}

In the case of non-zero PSF a source 
in some pixel may contribute gamma-rays to nearby pixels as well.
In order to correctly estimate the number of point sources
of a certain strength,
we need to know how often a gamma-ray from a source in one pixel
is detected in a different pixel.

\begin{figure}[t] %[htbp] here, top, bottom, page
\begin{center}
%\vspace{-12mm}
%\hspace{-3mm}
\epsfig{figure = 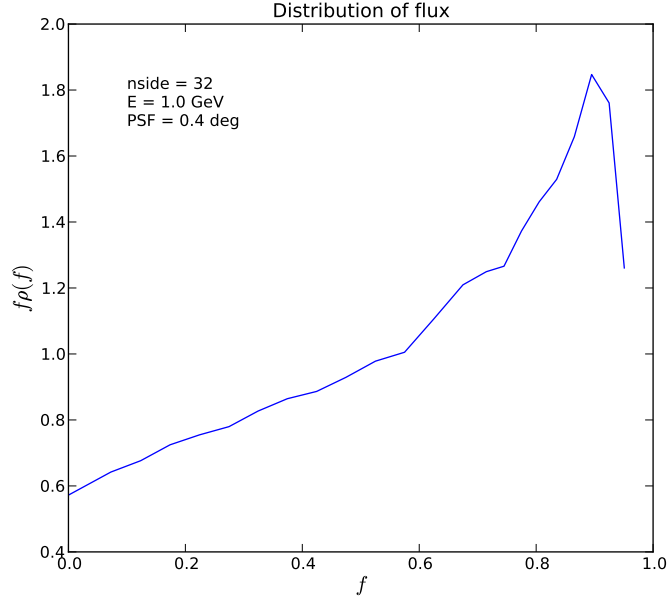, scale=\picsize} 
\end{center}
\vspace{-8mm}
\noindent
\caption{\small 
Average distribution of flux among pixels
from Gaussian sources at random positions.
The width of the Gaussian corresponds to the average point spread function
for front-converted gamma-rays with energies between 1 GeV and 300 GeV.
The pixels are determined in HEALPix \citep{2005ApJ...622..759G}
with ${\rm nside = 32}$.
%which approximately corresponds to pixel size of 2 degrees.
$x$-axis: fraction of the flux, 
$y$-axis: density of pixels as a function of $f$ (cf., Equation (\ref{eq:rho_def}))
multiplied by the flux fraction.
The fractions are computed by taking 50,000 Gaussian distributions 
at random positions on the sphere.
The area under the curve corresponds to the total flux equal to 1.
}
\label{fig:frhof}
\vspace{1mm}
\end{figure}

We can represent the effect of the PSF as a smearing of a point source over some area,
so that the flux from the point source is split between several pixels.
We determine the average properties of the flux splitting from Monte Carlo simulations.
In the simulation, we assume that the profile of the PSF is Gaussian.
In order to calculate the average splitting,
we put $n$ Gaussians at random positions on the sphere and
integrate every Gaussian over the pixels to get a collection
of fractions $f_i,\; i = 1, \ldots, N_{\rm pix}$, such that $f_1 + f_2 + \cdots = 1$
(in practice, one can truncate at some minimal value of $f$).
Denote by $\Dl n(f) $ the number of fractions for $n$ Gaussians that fall within
$\Dl f$, then the average distribution of fractions is defined as
\be
\lb{eq:rho_def}
\rho(f) = \left.
\frac{\Dl n(f)}{n \Dl f}\right|_{\Dl f \ra 0,\: n \ra \infty}.
\ee
This distribution is normalized as
\be
\lb{eq:norm-frac}
\int f \rho(f) df = 1.
\ee
The case of zero PSF corresponds to $\rho(f) = \dl(f - 1)$.
Sums over fractions are substituted by the integral $\int \rho(f) df$.
In particular, the contribution of point sources with intrinsic flux between $S$ and $S + dS$
to the total number of $m$-photon sources is
\be
d x_m = d N(S) \int df \rho(f) \frac{(fS)^m}{m!} e^{-fS}.
\ee
The expected number of  $m$-photon sources 
inside a pixel
including the effect of PSF is
\be
\lb{eq:xm_noPSF}
x_m = \frac{\Om_{\rm pix}}{4\pi} \int_0^\infty dS \frac{dN}{dS}(S) 
\int df \rho(f) \frac{(fS)^m}{m!} e^{-fS}.
\ee
An example of the function $\rho(f)$ relevant for the data analysis 
in Section \ref{sect:data}
is presented in Figure \ref{fig:frhof}.
If we rescale the integration variable $S$ in Equation (\ref{eq:xm_noPSF}) by $f$, then
the effect of the PSF 
can be represented as a change of the source-count function
\be
\lb{eq:new_source}
\frac{dN}{dS}(S)  \lra \frac{d\td N}{dS}(S) = \int df  \frac{\rho(f)}{f} \frac{dN}{dS}({S}/{f}).
\ee
%This new source count distribution,
%$\frac{d\td N}{dS}(S)$,
%takes into account the flux from a source that falls within one pixel.
The apparent number of sources 
with small fluxes in $\frac{d\td N}{dS}$ 
is larger than the corresponding number of physical sources
in $\frac{d N}{dS}$
due to contribution from point sources in nearby pixels,
whereas the apparent number of sources with large fluxes is smaller
due to loss of the flux to nearby pixels.
One can see the effect of PSF in Figure \ref{fig:datafit} on the right.
On the same figure, we also plot the expected number of $m$-photon sources for discrete $m$,
derived from $\frac{d\td N}{dS}$.

\subsection{The Model}
\lb{sect:sources}

In Equation (\ref{eq:genf}), the probabilities $p_k$ and the average expected number
of $m$-photon sources may depend on the position of the pixel.
The quantity that we use is the averaged probability $p_k = \frac{n_k}{N_{\rm pix}}$,
where $n_k$ is the number of pixels with $k$ photons.
The generating function for averaged probabilities is given by
\be
\lb{eq:genf-diff}
\sum_{k=0}^\infty {p}_k t^k = \frac{1}{N_{\rm pix}}
\sum_{{\rm p} = 1}^{N_{\rm pix}} {\rm exp}
\left(\sum_{m = 1}^\infty (x_m^{\rm p} t^m - x_m^{\rm p})\right).
\ee
This form of the generating function can be used when
the distribution of sources is not isotropic or the exposure is 
non-uniform.
If the exposure is sufficiently uniform, then
we may assume that the distribution of extragalactic sources is independent
of the pixel position and can be taken out of the sum over pixels.
The diffuse emission corresponds to ``one-photon'' source density
that depends on the pixel, $x_{\rm diff}({\rm p})$.
Thus, the expression on the right-hand side of Equation (\ref{eq:genf-diff})
can be split into a product 
\be
\frac{1}{N_{\rm pix}} \sum_{{\rm p} = 1}^{N_{\rm pix}} e^{x_{\rm diff}^{\rm p} t - x_{\rm diff}^{\rm p}
+{\sum_{m = 1}^\infty (x_m t^m - x_m)}}
=
\left(
\frac{1}{N_{\rm pix}} \sum_{{\rm p} = 1}^{N_{\rm pix}} e^{x_{\rm diff}^{\rm p} t - x_{\rm diff}^{\rm p}}
\right) \cdot
e^{\sum_{m = 1}^\infty (x_m t^m - x_m)}
\ee
The second term in the product on the right hand side 
corresponds to the isotropic distribution of AGN-like point sources
\be
\lb{eq:AGN-like}
P(t) = {\rm exp} \left(\sum_{m = 1}^\infty (x_m t^m - x_m)\right),
\ee
while the first term corresponds to diffuse emission or, indistinguishably, one-photon sources
\be
\lb{eq:diff}
D(t) = \frac{1}{N_{\rm pix}}
\sum_{{\rm p} = 1}^{N_{\rm pix}} e^{x_{\rm diff}^{\rm p} t - x_{\rm diff}^{\rm p}}.
\ee
It is convenient to separate the diffuse emission into a non-isotropic part
that puts a lower limit on Galactic diffuse emission and 
an isotropic part that consists of isotropic Galactic component and
a possible population of weak extragalactic sources
(additional to AGN-like sources)
\be
x_{\rm diff}^{\rm p} = x_{\rm Gal}^{\rm p} + x_{\rm isotr}.
\ee
The main problem with the non-isotropic diffuse emission 
is a variability on large scales
(on small scales the variability in Galactic emission is subdominant
to Poisson noise and variability due to the presence of undetected 
point sources).
In Appendix \ref{sect:LS} we construct a model for the 
non-isotropic component $x_{\rm Gal}^{\rm p}$
that varies on large scales.
We mask the known point sources and use
low multipole spherical harmonics in order to 
filter out the small-scale variations while preserving the large-scale
ones.
We add a constant to $x_{\rm Gal}^{\rm p}$ so that
${\rm min}(x_{\rm Gal}^{\rm p}) = 0$.
In the following, the non-isotropic component of 
diffuse emission is fixed.
The corresponding generating function is
\be
\lb{eq:ls_model}
G(t) = \frac{1}{N_{\rm pix}}
\sum_{{\rm p} = 1}^{N_{\rm pix}} e^{x_{\rm Gal}^{\rm p} t - x_{\rm Gal}^{\rm p}}.
\ee
We denote the probability generating function for the isotropic emission as
\be
I(t) = e^{x_{\rm isotr} t - x_{\rm isotr}}.
\ee

The total generating function for the probability distribution of photons in pixel is 
the product of the three components
\be
\lb{eq:prod_model}
\sum_{k=0}^\infty p_k t^k = P(t) \cdot G(t) \cdot I(t).
\ee
The parameters of the point-source distribution in Equation (\ref{eq:dNdS_power})
and the level of the isotropic flux $x_{\rm isotr}$ are found from fitting the 
probability distribution given in Equation (\ref{eq:prod_model})
to the observed probability distribution.
The details of the fitting algorithm are presented in the next subsection.

This simple factorized form of the generating function is valid in every pixel.
In general, this factorization is not possible if we take an average over pixels,
unless at most one sources is non-isotropic and
the other sources can be taken out of the average over pixels.
%but if we average over pixels to obtain the values of the probabilities $p_k$,
%then the factorization is not valid unless all but one source are isotropic.
%In our case, %in order to have this simple factorized form, 
%we need to assume a homogeneous exposure.
The factorization leads to a significant simplification of calculations
at the expense that, due to a variation in exposure,
a part of isotropic EGB may be misinterpreted
as a non-isotropic Galactic foreground.
This effect is expected to be small, provided that 
the \Fermi -LAT exposure is sufficiently uniform \citep{2009ApJ...697.1071A}.
Another possible source of systematic uncertainty is that the distribution of galaxies
in the local neighborhood of the Milky Way is non-uniform and a part of emission from these
galaxies can be misinterpreted as Galactic emission.

\subsection{\lb{sect:alg} Fitting Algorithm}

The algorithm has two main parts:
\begin{enumerate}
\item
Determine the non-isotropic part $x_{\rm Gal}^{\rm p}$ of Galactic diffuse emission.
%Variable part of Galactic diffuse emission:
We mask the bright point sources
and use low multipoles of spherical harmonics to
find the variable part of the remaining flux (Appendix \ref{sect:LS}).
Then we add a constant such that ${\rm min}(x_{\rm Gal}^{\rm p})  = 0$.
In the following  $x_{\rm Gal}^{\rm p}$ is fixed, i.e. we do not vary this component
in fitting the pixel counts.
\item
Use pixel counts to find $x_{\rm isotr}$ and $dN/dS$.
The source-count distribution has four parameters:
normalization, break, and two indices.
Together with the level of isotropic contribution,
$x_{\rm isotr}$, this gives five fitting parameters which we denote by 
$\al = (A, S_{\rm break}, n_1, n_2, x_{\rm isotr})$,
where $A$ denotes the normalization in the source-count distribution in
Equation (\ref{eq:dNdS_power}).
\end{enumerate}
The model prediction for the pixel counts $\nu_k(\al)$ is determined by multiplying the 
right-hand side of 
Equation (\ref{eq:prod_model}) by $N_{\rm pix}$
\be
\lb{eq:genf2}
N_{\rm pix}\cdot P(t) \cdot G(t) \cdot I(t)
= \sum_{k = 0} \nu_k(\al) t^k.
\ee
The expected pixel counts $\nu_k(\al)$ are found 
by expanding the left-hand side of this equation in powers of $t$
and picking the coefficient in front of $t^k$.
Given the observed number of pixels $n_k$ with $k$ photons,
the likelihood of $\nu_k(\al)$ is estimated by the Poisson probability 
\be
L_k = \frac{\nu_k^{n_k}}{n_k!} e^{-\nu_k}.
\ee
The overall likelihood of the model is estimated as
\be
\lb{eq:Lhood}
L(\al) = \prod_k \frac{\nu_k(\al)^{n_k}}{n_k!} e^{-\nu_k(\al)}.
\ee
Here the product is over the number of photons $k$,
while the data are represented by the number of pixels $n_k$
that contain $k$ photons.
This is different from a usual coordinate space fit of maps,
where the product is over pixels and the data are the number 
of photons $k_p$ in a pixel $p$.

We use this likelihood function to find the best-fit parameters $\al_*$ for a
given set of observed pixel counts $n_k$.
The significance of deviation of $\al$ from $\al_*$ can be estimated as
\be
\lb{eq:sigma_def}
\frac{\sm(\al)^2}{2} = \ln L(\al_*) - \ln L(\al).
\ee
In the case of large counts the likelihood is well approximated by the Gaussian distribution and
$\sm(\al)$ is the deviation in sigma values.

\section{Data analysis}
\lb{sect:data}

\begin{figure}[t] %[htbp] here, top, bottom, page
\begin{center}
%\vspace{-12mm}
\hspace{-3mm}
\epsfig{figure = 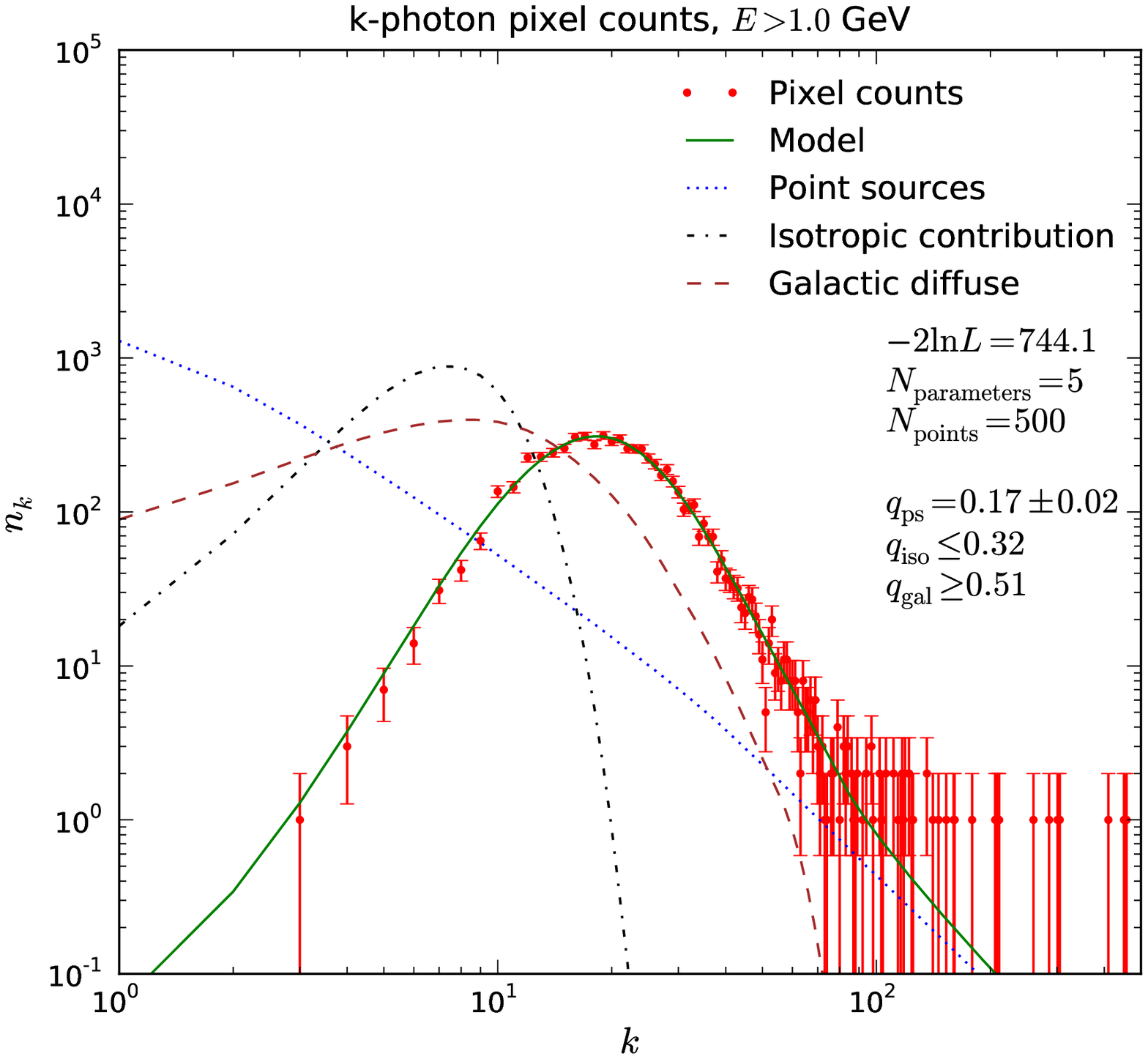, scale=\picsizetwo} 
\hspace{-3mm}
\epsfig{figure = 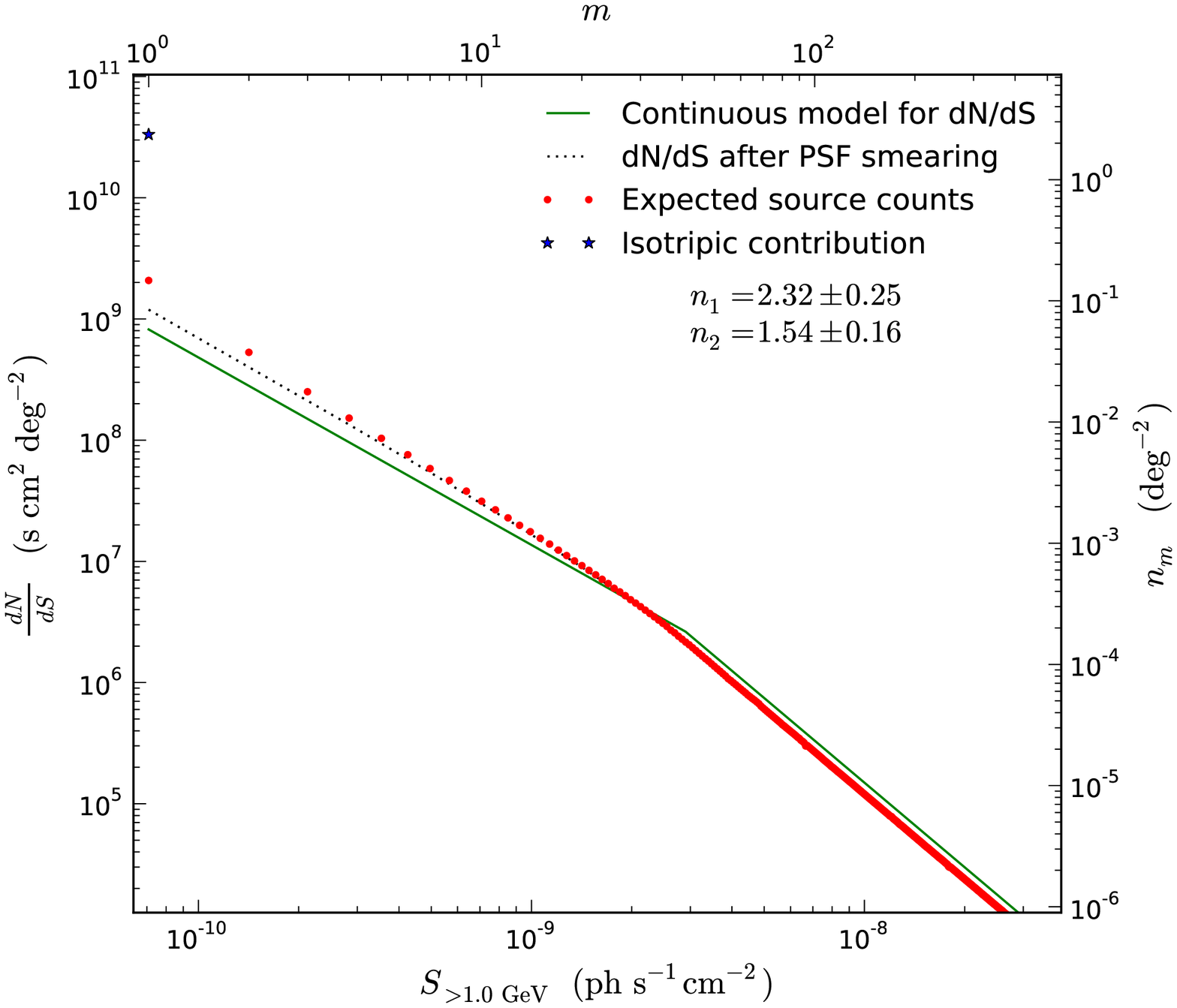, scale=\picsizetwo}
\end{center}
\vspace{-8mm}
\noindent
\caption{\small 
Left plot: $n_k$ is the number of pixels with $k$ photons,
the red dots correspond to pixel counts derived from \Fermi data,
the errors bars are equal to $\sqrt{n_k}$.
We consider three sources: AGN-like point sources (blue dotted line),
isotropic Poisson contribution (brown dashed line),
and non-isotropic Galactic diffuse emission (black dash-dotted line).
Note, that the total model on this plot is not the sum of the components,
the corresponding generating function of the PDF is
a product of the generating functions in Equation (\ref{eq:prod_model}).
$N_{\rm points}$ corresponds to the number of points on the $x$-axis that we have used for fitting,
$k < 500$.
Right plot: green solid line is the physical model for the AGN-like source counts,
blue dotted line is the source counts in pixels in the presence of PSF 
(Equation (\ref{eq:new_source})),
points represent expected number of $m$-photon sources per ${\rm deg^2}$
as defined in Equation (\ref{eq:xm_noPSF}),
blue star is the value of the isotropic flux.
}
\label{fig:datafit}
\vspace{1mm}
\end{figure}

We consider 11 months of \textsl{Fermi} data (August 4, 2008 - July 4, 2009)
that were used for the first \textsl{Fermi} point-source catalog
\citep{2010ApJS..188..405A}
and for the \Fermi analysis of the source-count distribution
\citep{2010ApJ...720..435A}.

In order to reduce the PSF, we use front-converted gamma-rays, i.e.,
the gamma-rays converted in the front part of the \textsl{Fermi}-LAT tracker
\citep[][]{2009ApJ...697.1071A},
with energies between 1 GeV and 300 GeV.
The corresponding (quadratic) average of the PSF is $0.4^\circ$.%
\footnote{
\url{http://www-glast.slac.stanford.edu/software/IS/glast_lat_performance.htm}
}
For pixelation of data, we use HEALPix \citep{2005ApJ...622..759G}
with the pixelation parameter nside = 32,
which corresponds to pixel size about $2$ degrees.%
\footnote{
We consider the cases of nside = 16 and nside = 64 in Appendix
\ref{sect:PIX}.}
We mask the galactic plane within 30 degrees in latitude. %, $-30^\circ < b < 30^\circ$.
The number of pixels outside of the mask is $N_{\rm pix} = 6178$ and 
the number of photons is 152,143.

The distribution of pixel counts is presented in Figure \ref{fig:datafit} on the left.
%where $n_k$ is the number of pixels that contain $k$ photons.
%The corresponding probability distribution is $p_k = n_k / N_{\rm pix}$.
%In order to fit the observed pixel counts,
We model this distribution by a combination of 
three components: non-isotropic Galactic diffuse emission derived in Appendix \ref{sect:LS},
isotropic distribution of photons, and the distribution of AGN-like point sources
described in Section \ref{sect:sources}.
The best-fit models for the isotropic flux and the AGN-like point sources are
presented in Figure \ref{fig:datafit} on the right.
%Every component corresponds to a generating function and
%we show the pixel counts corresponding to the three components
%on the left plot of Figure \ref{fig:datafit}.
%the total model is obtained by multiplication of the corresponding generating functions
%(cf., Equation \ref{eq:prod_model}).

\begin{figure}[t] %[htbp] here, top, bottom, page
\begin{center}
\vspace{-5mm}
%\hspace{-3mm}
\epsfig{figure = 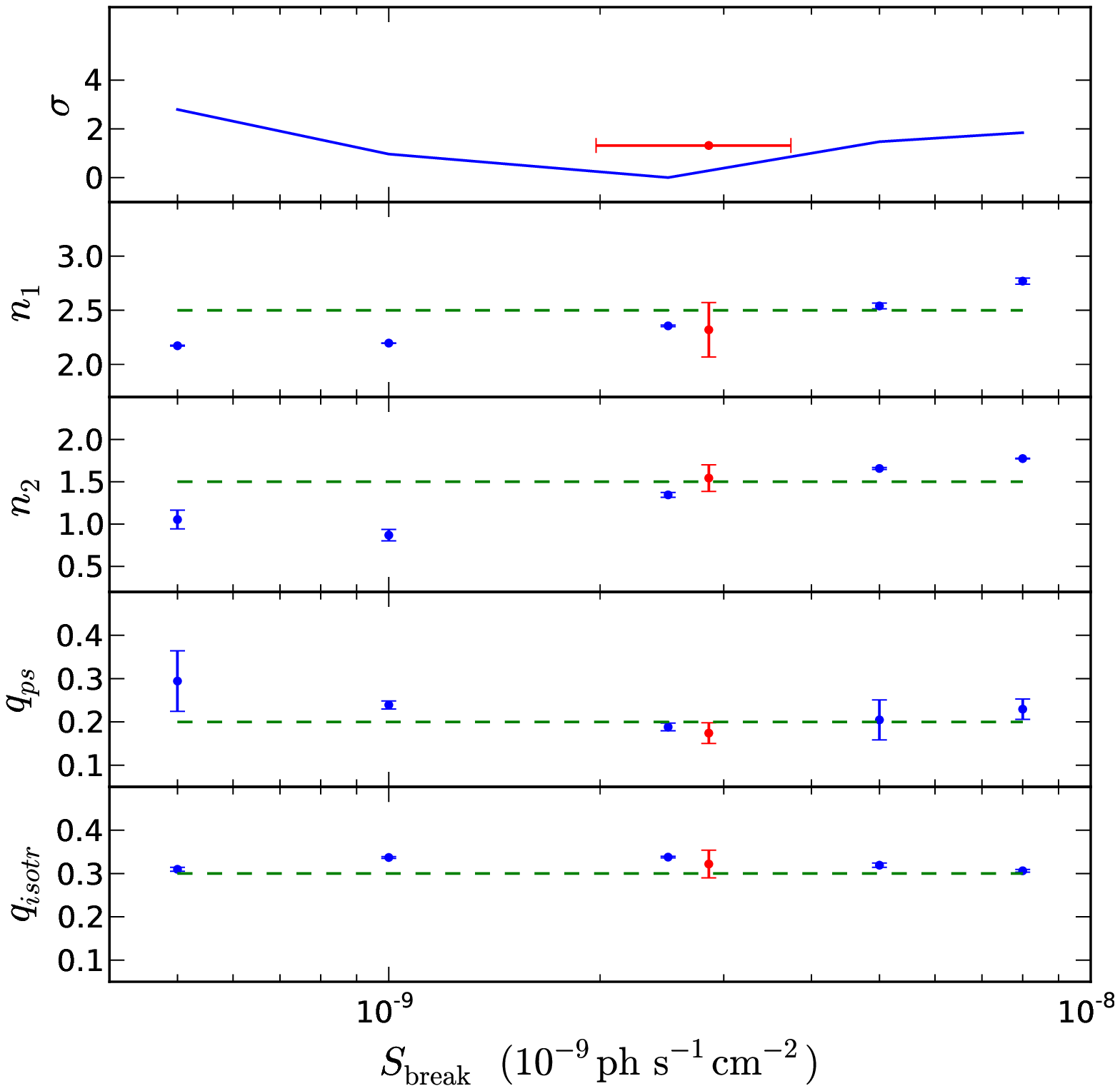, scale=\picsize} 
\end{center}
\vspace{-8mm}
\noindent
\caption{\small 
Results of fitting the AGN-like source-count distribution for fixed positions of the break.
The significance of deviation is determined from Equation (\ref{eq:sigma_def})
as $\sm = \sqrt{2\Dl\ln L}$.
$n_1$ and $n_2$ are the indices above and below the break for differential source-count distribution,
$q_{\rm ps}$ is the fractional contribution from AGN-like point sources,
$q_{\rm isotr}$ is the fractional contribution from an isotropic source
(it provides an upper limit on the extra-galactic isotropic contribution
additional to AGN-like point sources).
Red dots correspond to the model used in Figure \ref{fig:datafit}.
Green dashed lines provide reference values:
$n_1 = 2.5$, $n_2 = 1.5$, $q_{\rm ps} = 0.2$, and $q_{\rm isotr} = 0.3$.
}
\label{fig:scan}
\vspace{1mm}
\end{figure}

In fitting we consider pixel counts $n_k$ with $k < 500$,
i.e., the number of data points is $N_{\rm points} = 500$.
We use $\ell_{\rm max} = 20$ for the Galactic diffuse emission model.
%The model is described in Section \ref{sect:model} .
The sampling is performed with Metropolis-Hastings 
Markov Chain Monte Carlo method with 1500 steps.
The average model is presented in Figure \ref{fig:datafit}
and compared with \textsl{Fermi} models in Table \ref{tab:compare}.
We find that the position of the break is almost a flat direction.
The results of fitting with a set of fixed break positions are presented in Figure \ref{fig:scan}.

\begin{table*}[h]%[tpb]
\begin{center}
\caption{\label{tab:compare}\small
Comparison with \textsl{Fermi} Results \citep{2010ApJ...720..435A}.
}
\vspace{3mm}
\footnotesize
\begin{tabular} {|cccc|}
\hline
Analysis & $n_1$ & $n_2$ & $S_{\rm break}^a$\\
\hline
This work & $2.31\pm 0.25$ & $1.54\pm 0.16$ & $2.9\pm 0.9$ \\
\Fermi $>100$ MeV & $2.44\pm 0.11$ & $1.58\pm 0.08$ & $2.8 \pm 0.5^b$  \\
\Fermi 1 - 10 GeV & $2.38^{+0.15}_{-0.14}$ & $1.52^{+0.8}_{-1.1}$ 
			& $\rm 2.3 \pm 0.6 $  \\
\hline 
\end{tabular}
\end{center}
{\footnotesize 
$^a$ In units of   $10^{-9}{\rm ph\: s^{-1} cm^{-2}}$.\\
$^b$ Rescaled from the break in \cite{2010ApJ...720..435A} 
according to energy spectrum $\sim E^{-2.4}$.
}
\end{table*}

We find that the contribution of point sources to the gamma-ray flux is 
$(17 \pm 2)\%$.
This fraction is obtained by integrating the best-fit source-count model presented in
Figure \ref{fig:datafit} and in Table \ref{tab:compare} from zero to infinity.
This fraction is not model independent, i.e.,
it may change for different shapes of the source-count function used in fitting,
but, provided that the broken power-law source-counts distribution
gives a good fit to the data,
we do not expect a large systematic uncertainty due to the change in the 
shape of the fitting function.
In part this can be justified by looking at the models with fixed positions of the break:
most of the models in Figure \ref{fig:scan}
%the models with small deviation from the best fit model 
have the point-source fraction $q_{\rm ps} \lesssim 0.25$.

At smaller position of the break the contribution of point sources becomes larger
($q_{\rm ps} \sim 0.3$ for the left point in Figure \ref{fig:scan})
but the fit has smaller likelihood.
It would be interesting to estimate constraints on models
involving high redshift AGNs to explain the unresolved part of the EGB
\citep[e.g.,][]{2010arXiv1012.1247A, 2011arXiv1103.3484N}
using the statistics of photon counts.

We can also put an approximate lower bound of 51\% on Galactic diffuse emission,
and an approximate upper bound of 32\% on isotropic emission consistent with 
Poisson statistics.
%Some part of this isotropic emission can be attributed to star-forming galaxies \citep{2010ApJ...722L.199F}.
In terms of the absolute flux values,
the total gamma-ray flux above 1 GeV for $|b| > 30^\circ$
is $F_{\rm tot} = 1.75\times 10^{-6}\: {\rm s^{-1} cm^{-2} sr^{-1}}$,
the diffuse Galactic flux is 
$F_{\rm Gal} > 8.8\times 10^{-7}\: {\rm s^{-1} cm^{-2} sr^{-1}}$,
the isotropic component of the flux
is 
$F_{\rm isotr} = (5.6 \pm 0.6) \times 10^{-7}\: {\rm s^{-1} cm^{-2} sr^{-1}}$,
the flux from AGN-like point sources with $dN/dS$ parameters given in 
Table \ref{tab:compare} and in Figure \ref{fig:datafit} is
$F_{\rm PS} = (3.0 \pm 0.4) \times 10^{-7}\: {\rm s^{-1} cm^{-2} sr^{-1}}$.

\begin{figure}[t] %[htbp] here, top, bottom, page
\begin{center}
%\vspace{-12mm}
\hspace{-3mm}
\epsfig{figure = 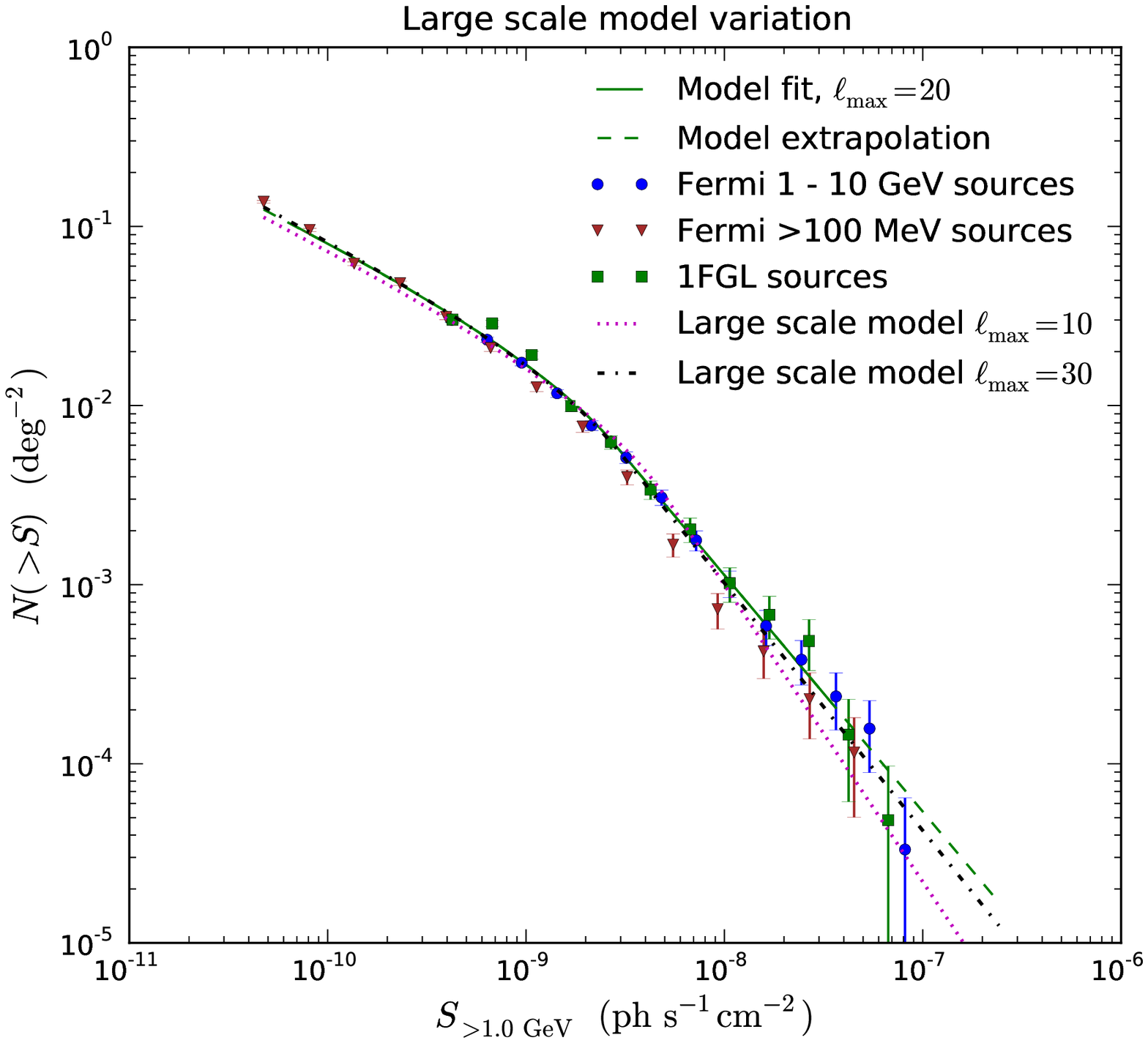, scale=\picsizetwo} 
\hspace{-3mm}
\epsfig{figure = 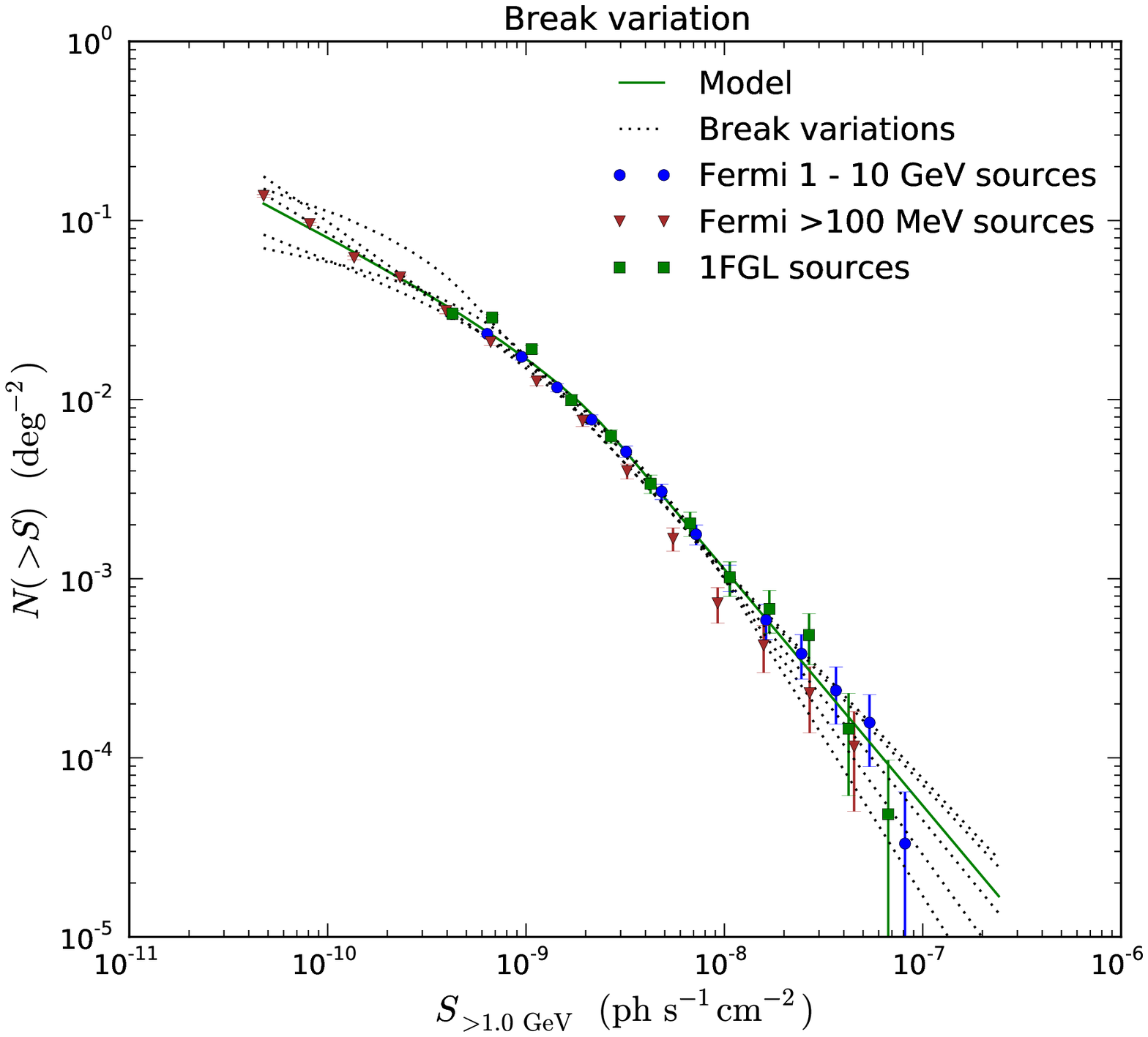, scale=\picsizetwo}
\end{center}
\vspace{-8mm}
\noindent
\caption{\small 
Left plot: different lines correspond to different models for Galactic diffuse emission of gamma-rays
corresponding to $\ell_{\rm max} = 10,\: 20,\: 30$
%In the models we use small $\ell$ spherical harmonics of the data with known point
%sources subtracted to model the large distribution of Galactic diffuse emission
(Appendix \ref{sect:LS}).
Solid green line corresponds to the range of photon counts (1 to 500) used in fitting.
Dashed green line is the extrapolation of the model.
Magenta triangles correspond to total source counts found by \textsl{Fermi} collaboration
above 100 MeV \citep[right plot in Figure 9 of][]{2010ApJ...720..435A} with
flux rescaled assuming energy density spectrum $\sim E^{-2.4}$.
Blue circles correspond to sources found by the Fermi collaboration in
$1 - 10$ GeV energy bin (center plot in Figure 17 of \cite{2010ApJ...720..435A}).
Green squares correspond to $1 - 100$ GeV flux for sources in the first 
\Fermi catalog \citep{2010ApJS..188..405A}.
Right plot: models with different fixed positions of the break presented in Figure \ref{fig:scan}.
}
\label{fig:Fermi}
\vspace{1mm}
\end{figure}

Using this value of the total flux from AGN-like point sources,
we can estimate the gamma-ray flux from undetected sources.
The total flux in the energy range 1 - 100 GeV
from point sources at $|b| > 30^\circ$ detected by \textsl{Fermi} \citep{2010ApJS..188..405A} is
$F_{\Fermi\: \rm AGN} = 2.1 \times 10^{-7}\:  {\rm s^{-1} cm^{-2} sr^{-1}}$.
Consequently, the flux from undetected AGN-like point sources above $30^\circ$ can be estimated as
$F_{\rm unres\:  AGN} = 0.9 \times 10^{-7}\:  {\rm s^{-1} cm^{-2} sr^{-1}}$.
This value is reasonable, since from Figure \ref{fig:Fermi} it follows that the detected point sources
give a good approximation to the best-fit distribution of sources down to fluxes below the break in $N(S)$,
while the contribution of point sources to the gamma-ray background is saturated around the break.
The extragalactic diffuse emission can be estimated from
the model presented in \cite{2010PhRvL.104j1101A}.
In this model, the EGB flux above 1 GeV is
$F_{\rm EGB} \approx 4 \times 10^{-7}\:  {\rm s^{-1} cm^{-2} sr^{-1}}$.
The contribution of unresolved AGN's to this EGB flux is about $23\%$.
This value is consistent with the estimations in \cite{2010ApJ...720..435A}.
We would like to stress that our analysis is different from 
\cite{2010ApJ...720..435A}.
We do not count any sources, but rather do a ``blind'' statistical analysis.
The consistency of results may serve as a non-trivial check of both methods.

\section{Discussion}
\lb{sect:discus}

One of the most interesting problems in gamma-ray astrophysics
is to understand the origin of EGB flux,
which is defined as the total gamma-ray flux minus the contribution from
resolved point sources, minus the Galactic foreground
\citep{2010PhRvL.104j1101A}.
In general, we can separate the sources of diffuse EGB
into two big classes: non-Poisson sources and Poisson-like sources.
In the first case, the contribution comes from sources below detection limit 
that may emit several photons during the observation time,
these are the AGN-like point sources
\citep[e.g.,][]{1993MNRAS.260L..21P, 1995ApJ...452..156C, 1996ApJ...464..600S,
2010arXiv1012.1247A, 2011arXiv1103.3484N}.
In this case the statistics of photon counts in pixels across the sky will be different from 
Poisson statistics due to correlation among photons coming from the same point source.
In the second case, there is a large number
of sources emitting on average much fewer than one photon during the observation time,
e.g., star-forming and star-burst galaxies 
\citep{2002ApJ...575L...5P, 2007ApJ...654..219T, 2010ApJ...722L.199F}.
The statistics of photons in this case will be very close to Poisson statistics.

These two possibilities are relatively hard to separate based only on energy spectrum arguments
\citep[see, however,][]{2010arXiv1012.1247A, 2010arXiv1012.3678S, 2011arXiv1103.3484N}.
Alternatively, one can use a correlation between gamma-ray emission and
multi-wavelength emission from AGNs
\citep{1995PASP..107..803U, 2000MNRAS.312..177M, 
2010ApJ...716...30A, 2010arXiv1006.5610A, 2011arXiv1103.4545L}
together with AGN population studies % at various frequencies
\citep[e.g.,][]{1995ApJ...444..567P, 2007ApJ...654..731H, 2007ApJ...662..182P, 2008MNRAS.385..310R, 
2009AJ....138..900H, 2009ApJ...696...24S}
to infer the AGN contribution to gamma-ray background.

In this paper we show that the statistics of photons in pixels can also be used to
constrain the distribution of point sources below the detection limit.
We find that there is a slight preference for two populations of point sources:
the AGN-like point sources with a break at relatively high flux 
%\citep{2010ApJ...720..435A}
and a population of faint sources such as star-forming galaxies
rather than a single population of AGN-like point sources with a break at a smaller flux.
This statement is insensitive to the energy spectrum since we use the integrated
flux above 1 GeV but it may depend on the form of the function used to fit 
the point source number counts.
Possible sources of systematic uncertainty 
include the source smearing due to PSF,
non-isotropic Galactic diffuse emission, 
non-isotropic emission (on large scales) from nearby galaxies,
and non-homogeneous exposure.
The interpretation of the point-source number counts may also be affected
by clustering of sources on scales smaller than the PSF.

A stronger constrain on the population of gamma-ray point sources may come from
an unpixelized analysis of the data.
In this more general analysis, the likelihood of a model depends on the full gamma-ray data 
(positions on the sky, energy, arrival time)
rather than the counts of gamma-rays in pixels.
This likelihood would have the form of an integral over flux times the 
source number counts, times an integral over all possible positions of the sources on the sky.
The photon counts approximation is simpler computationally and 
already gives rough constraints on the source population,
while the full unbinned analysis may provide stronger constraints 
at the expense of computational complexity.
Examples of unbinned analysis using some part of gamma-ray data
include the gamma-ray two-point correlation function 
\citep[e.g.,][]{2009JCAP...07..023A,2010arXiv1012.1873G},
angular power spectrum
\citep{2006PhRvD..73b3521A, 2010ApJ...723..277H},
nearest neighbor statistics
\citep{2011arXiv1101.0256S},
etc.

We also believe that the techniques of generating 
functions in the study of photon counts statistics,
known for some time in radio and X-ray observations,
may have further important applications in data analysis of
current and future radio, IR, X-ray, and gamma-ray observations.

{\large \bf Acknowledgments.}

\noindent
The authors are thankful to 
Kevork Abazajian,
Marko Ajello,
Jo Bovy,
Christopher Kochanek,
Jennifer Siegal-Gaskins,
Tracy Slatyer,
Tonia Venters,
and Itay Yavin
for stimulating discussions and useful comments.
We are also thankful to Ilias Cholis for the help in selecting the \textsl{Fermi} gamma-ray data.
This work is supported in part by the Russian Foundation of Basic Research under 
Grant No. RFBR 10-02-01315 (D.M.), by the NSF Grant No. PHY-0758032 (D.M.),
by Packard Fellowship  No. 1999-1462 (D.M.),
by NASA grant NNX08AJ48G (D.W.H.), and by the NSF grant AST-0908357 (D.W.H.).
Some results in this paper have been
obtained using the HEALPix package \citep{2005ApJ...622..759G}.

\appendix
\section{Derivation of statistics}
\lb{sect:stat}

In this Appendix we derive the photon counts statistics in the presence of point sources.
The derivation is based on the following well known property of generating functions for probabilities:
the generating function of a sum of two independent sources is the product of the corresponding
generating functions \citep[e.g.,][Section 3.6]{Hoel1971}.
Indeed, consider a box where we can put red balls with probabilities $p_k$, $k = 1,\:2,\:3\ldots$,
and blue balls with corresponding probabilities
$q_k$, $k = 1,\:2,\:3\ldots$,
then the probability to find $k$ balls of any color is
\be
r_k = \sum_{k' = 0}^k p_{k'} q_{k - k'}, \qquad k = 1,\:2,\:3\ldots
\ee
The last equation is the rule for multiplication of polynomials or power series.
Let
\be
R(t) = \sum_k r_k t^k,\qquad
P(t) = \sum_k p_k t^k,\qquad
Q(t) = \sum_k q_k t^k,
\ee
then
\be
R(t) = P(t) \cdot Q(t).
\ee
%We will associate a source that contributes $k$ gamma-rays in a pixel with probabilities
As before, denote the probability to observe $k$ photons in a pixel by $p_k$. 
The corresponding generating function is
\be
P(t) = \sum_k p_k t^k.
\ee
The knowledge of the generating function is equivalent to the knowledge of all $p_k$,
provided that every $p_k$ can be found from $P(t)$ by picking the 
term in front of $t^k$.

We will now derive the generating function for probabilities to observe $k$ photons from
a collection of point sources.
Denote by $x_m$ the average number of point sources inside a pixel that
emit exactly $m$ photons during the time of observation.
We assume that the probability to find $n_m$ $m$-photon sources in a pixel is 
given by the Poisson statistics
\be
p_{n_m} (x_m) = \frac{x_m^{n_m}}{n_m !} e^{-x_m},
\ee
then the probability to find $k$ photons from $m$-photon sources is
\be
p^{(m)}_k = 
\left\{
\ba{ll}
p_{n_m} (x_m), & \text{if $k = m \cdot n_m$ for some $n_m$}; \\
0, & \text{otherwise}.
\ea
\right.
\ee
The corresponding probability generating function is
\be
P^{(m)}(t) = \sum_k p_k t^k = 
\sum_{n_m} t^{m \cdot n_m} \frac{x_m^{n_m}}{n_m !} e^{-x_m} 
= e^{x_m t^m - x_m}.
\ee
The generating function of probabilities $p_k$ to observe $k$ photons from any sources 
is the product of the generating functions for every $m$
\bea
\nonumber
\sum_{k=0}^\infty p_k t^k &=& \prod_{m = 1}^\infty  e^{x_m t^m - x_m} \\ %{\rm exp} \left( x_m t^m - x_m \right) \\
\lb{eq:genf_app}
&=& {\rm exp}\left(\sum_{m = 1}^\infty (x_m t^m - x_m)\right).
\eea
If we denote $t = e^{2\pi i \om}$ and 
\bea
\td P(\om) &=& \sum_{k = 0}^\infty p_k e^{2\pi i \om k}, \\
\td X(\om) &=& \sum_{m = 0}^\infty x_m e^{2\pi i \om m},
\eea
then Equation (\ref{eq:genf_app}) can be represented as
\be
\td P(\om) = e^{\td X(\om) - \td X(0)},
\ee
which is the discrete Fourier transform analog of characteristic functions considered
in, e.g., %Equation (11) of 
\cite{1957PCPS...53..764S}.

\section{Non-isotropic Galactic diffuse emission}
\lb{sect:LS}

In this appendix, we describe a model for the non-isotropic Galactic diffuse emission.
This signal is ``filtered'' by low multipoles of the gamma-ray data.
%Our method allows one to find a part of the Galactic emission that varies over the sky.
In this approach, the homogeneous part of Galactic emission is indistinguishable from extragalactic flux.
Also some part of extragalactic emission from galaxies and galaxy clusters close to Milky Way
may lead to a signal that varies over the sky and can be misinterpreted as non-isotropic Galactic emission.
However, the majority of extragalactic emission comes from higher redshifts where 
the distribution of extragalactic sources is sufficiently uniform on large scales.

\begin{figure}[t] %[htbp] here, top, bottom, page
\begin{center}
\epsfig{figure = 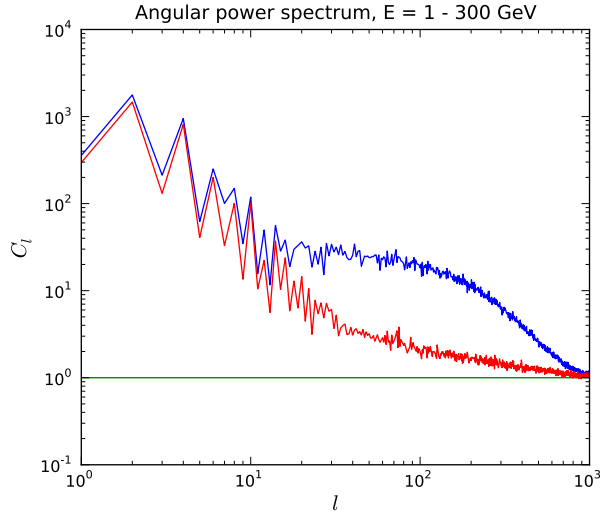, scale=\picsize}
\end{center}
\vspace{-8mm}
\noindent
\caption{\small 
%Angular power spectrum for photon counts at latitudes $|b| > 30$ deg.
Blue line (upper): angular power spectrum of gamma-ray data with $|b| > 30^\circ$.
Red line (lower): same as the blue line, but with additional masking of \textsl{Fermi} gamma-ray point sources
(Figure \ref{fig:maps}).
The normalization is chosen such that for the Poisson noise $\bra C_l \ket = 1$
(constant green line).
$C_l$'s below $\ell \sim 10$ are dominated by the variation in
Galactic diffuse emission on large scales.
Constant $C_l$'s above $\ell = 10$ are due to contribution of point sources.
Decay of $C_l$'s above $\ell \sim 100$ for the blue line is due to detector PSF.
In our case $\bra \rm PSF \ket \approx 0.4^\circ$, which corresponds to $\ell \gtrsim 400$.
A smoothed model for large-scale structure distribution is obtained by spherical 
harmonics decomposition up to $\ell = 20$,
the corresponding map of the model is presented in Figure \ref{fig:maps}.
}
\label{fig:Cls}
\vspace{1mm}
\end{figure}

\newcommand{\mapsize}{0.35}
\begin{figure}[t] %[htbp] here, top, bottom, page
\begin{center}
\vspace{-4mm}
\epsfig{figure = 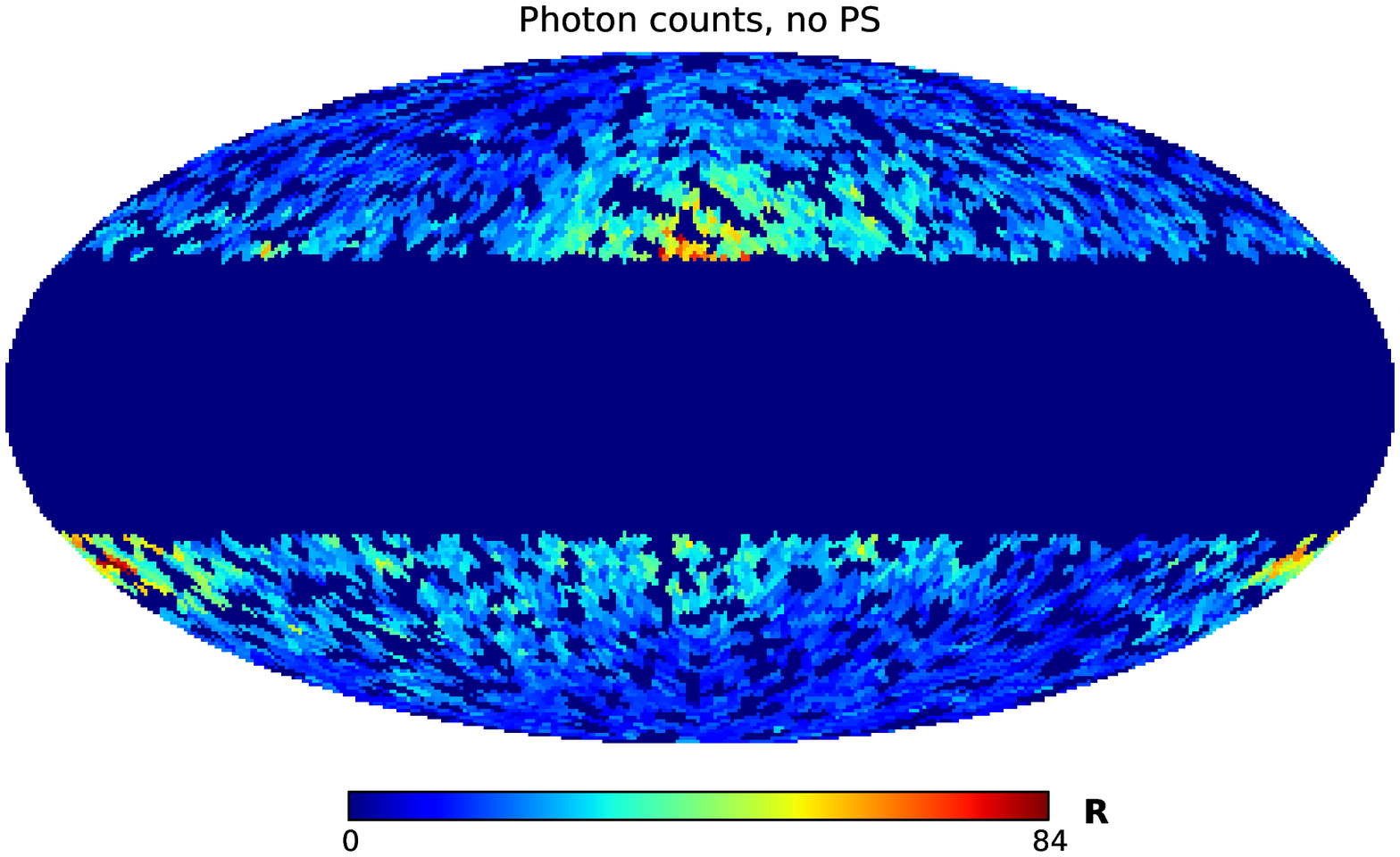, scale=\mapsize} \\
\vspace{-4mm}
\epsfig{figure = 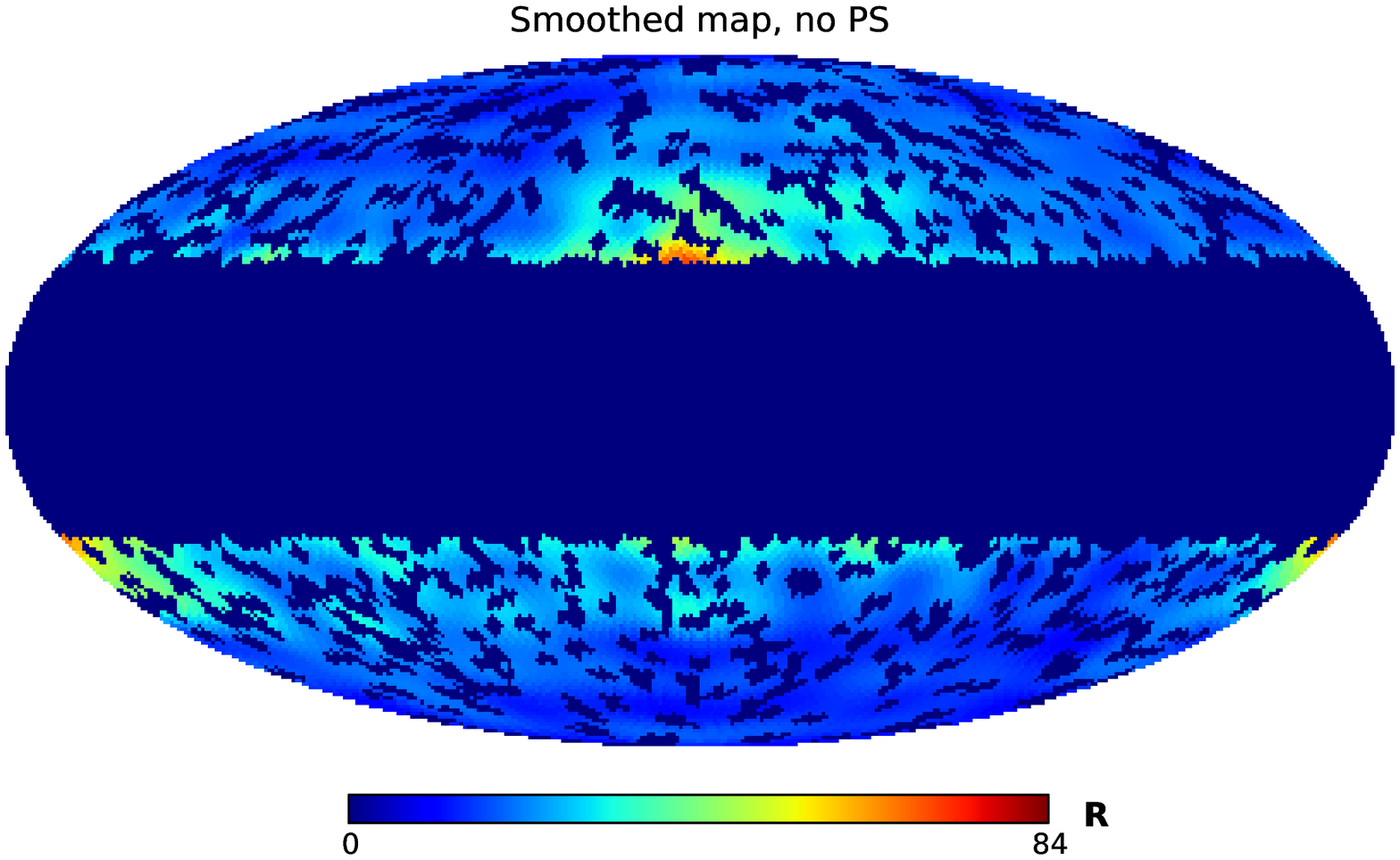, scale=\mapsize} \\
\vspace{-4mm}
\epsfig{figure = 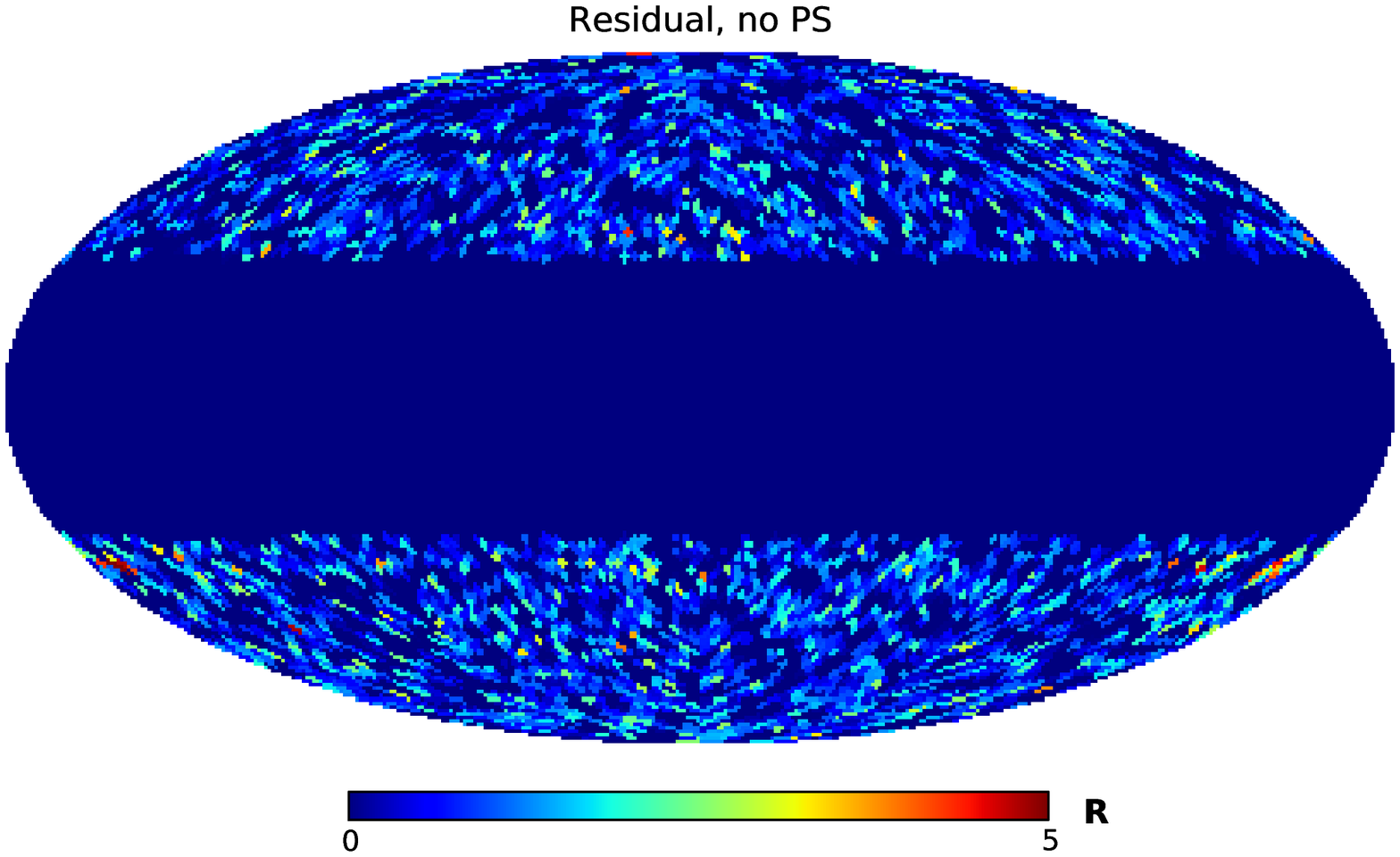, scale=\mapsize}
\end{center}
\vspace{-8mm}
\noindent
\caption{\small 
Sky maps for photon counts and for a Galactic diffuse emission model
derived from the photon counts by using 
the spherical harmonics decomposition with $l \leq 20$ (Equation (\ref{eq:x_diff})).
Residuals are in sigma values: $\frac{|{\rm model} - {\rm data}|}{\sqrt{{\rm model}}}$.
}
\label{fig:maps}
\vspace{1mm}
\end{figure}

In Figure \ref{fig:Cls}, we plot the angular power spectrum for the data at high latitudes ($|b| > 30^\circ$)
before and after masking the point sources in the first \textsl{Fermi} catalog 
\citep{2010ApJS..188..405A}.
The angular power spectrum is
\be
C_l = \frac{1}{2l + 1}\sum_m |a_{lm}|^2,
\ee
where $a_{lm}$'s are the spherical harmonics coefficients of $f(p)\cdot m(p)$, where
$f(p)$ is the number of photon counts inside pixel $p$ and
$m(p)$ is the mask function.
The mask function is equal to one (zero) for  $|b| > 30^\circ$ ($|b| < 30^\circ$).
In the case of masked point sources, we also have $m(p) = 0$
when a \textsl{Fermi} point source is inside the pixel or within two PSF from
the boundary of the pixel.
We also subtract the average of the 
data within the unmasked region in order to avoid a non-trivial
contribution from a constant source inside the window.
We choose the normalization such that for the Poisson noise $\bra C_l \ket = 1$.

\begin{figure}[t] %[htbp] here, top, bottom, page
\begin{center}
\epsfig{figure = 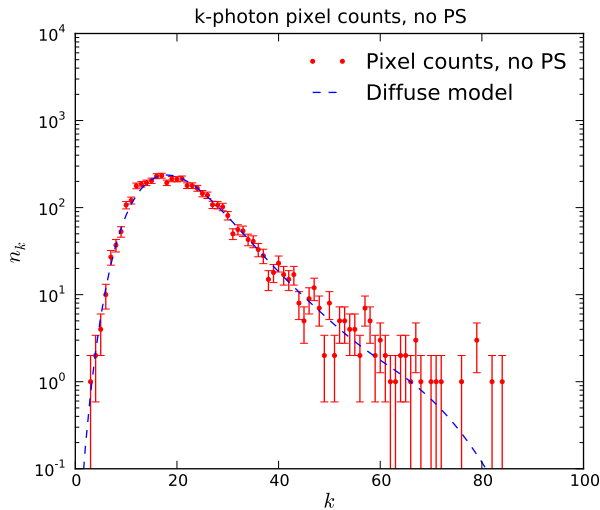, scale=\picsize}
\end{center}
\vspace{-8mm}
\noindent
\caption{\small 
Photon counts vs smooth model with \Fermi point sources subtracted.
The Galactic diffuse emission model is obtained by taking the spherical harmonics of the data
with $\ell \leq 20$.
The model photon counts are derived from the probability generating function in Equation
(\ref{eq:diff}) with $x_{\rm diff}$ given in Equation (\ref{eq:x_diff}).
}
\label{fig:model_ps}
\vspace{1mm}
\end{figure}

The algorithm for estimating the component that varies on large scales
is as follows:
\ben
\item
Calculate the $a_{lm}$'s inside a window larger than needed (in order to avoid edge effects).
We use $|b| > 20^\circ$ for the data above $30^\circ$.
We also fill in the pixels with point sources by the average of nearest neighbors.
\item
The large-scale distribution of gamma-rays is defined as
\be
\lb{eq:x_diff}
x_{\rm diff}(p) = A \sum_{l = 0}^{l_{\rm max}}\sum_{m = -l}^l a_{lm} Y_{lm}(p) + B,
\ee
where we find the coefficients $A$ and $B$ from the best fit of the diffuse model
to the photon counts in the pixels without point sources (see Figure \ref{fig:model_ps}).
\item 
We represent $B = B_{\rm min} + x_{\rm isotr}$ so that
\be
\lb{eq:x_gal}
x_{\rm Gal}(p) = A \sum_{l = 0}^{l_{\rm max}}\sum_{m = -l}^l a_{lm} Y_{lm}(p) + B_{\rm min}
\ee
is non-negative for $|b| > 30^\circ$.
In fitting to full data, $A$ and $B_{\rm min}$ are fixed, while $x_{\rm isotr}$ is allowed to vary
together with parameters describing the AGN-like point sources.
\een
An example of the diffuse emission model $x_{\rm diff}(p)$ for $l_{\rm max} = 20$
is presented in the middle plot of Figure \ref{fig:maps}.
The top plot represents the counts of photons in pixels inside the window $|b| > 30^\circ$
with masked \textsl{Fermi} gamma-ray point sources.
The bottom plot represents the deviation of the model from the data in ``sigma'' values.
In Figure \ref{fig:Fermi} we study the effect of changing $l_{\rm max} = 10, 20, 30$.
The difference is rather small, i.e., already $l_{\rm min} = 10$ captures the large-scale 
distribution of gamma-rays reasonably well.

\section{Variation of pixel size}
\lb{sect:PIX}

In the data analysis in Section \ref{sect:data}, 
we use the pixel size of about $2^\circ$ corresponding
to HEALPix parameter nside = 32.
In this appendix, we repeat the analysis of Section \ref{sect:data} for different sizes of pixels,
in particular we consider nside = 64 and nside = 16 corresponding to pixel sizes $1^\circ$ and $4^\circ$
respectively.

\begin{figure}[t] %[htbp] here, top, bottom, page
\begin{center}
\epsfig{figure = 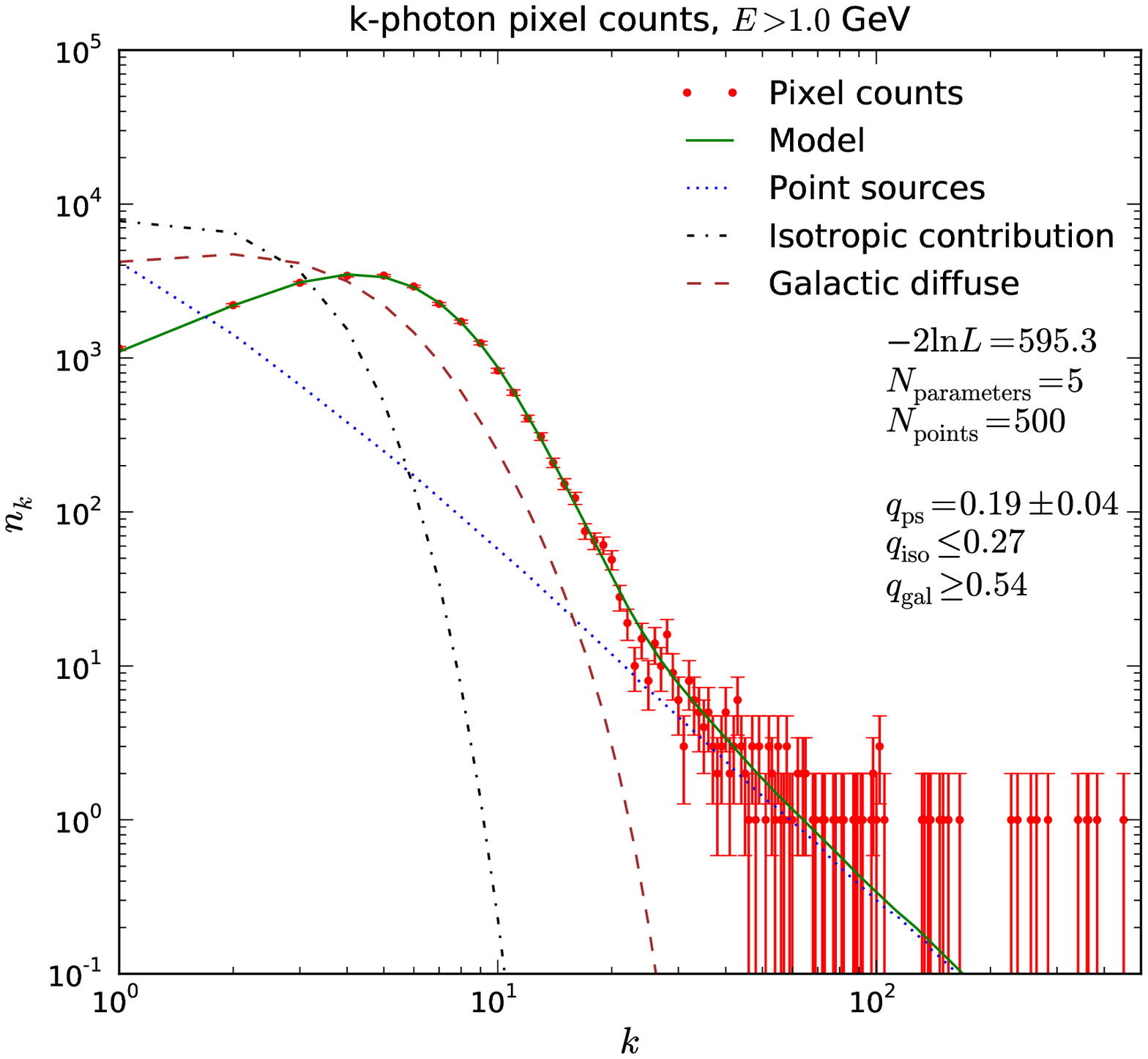, scale=\picsizetwo} 
\epsfig{figure = 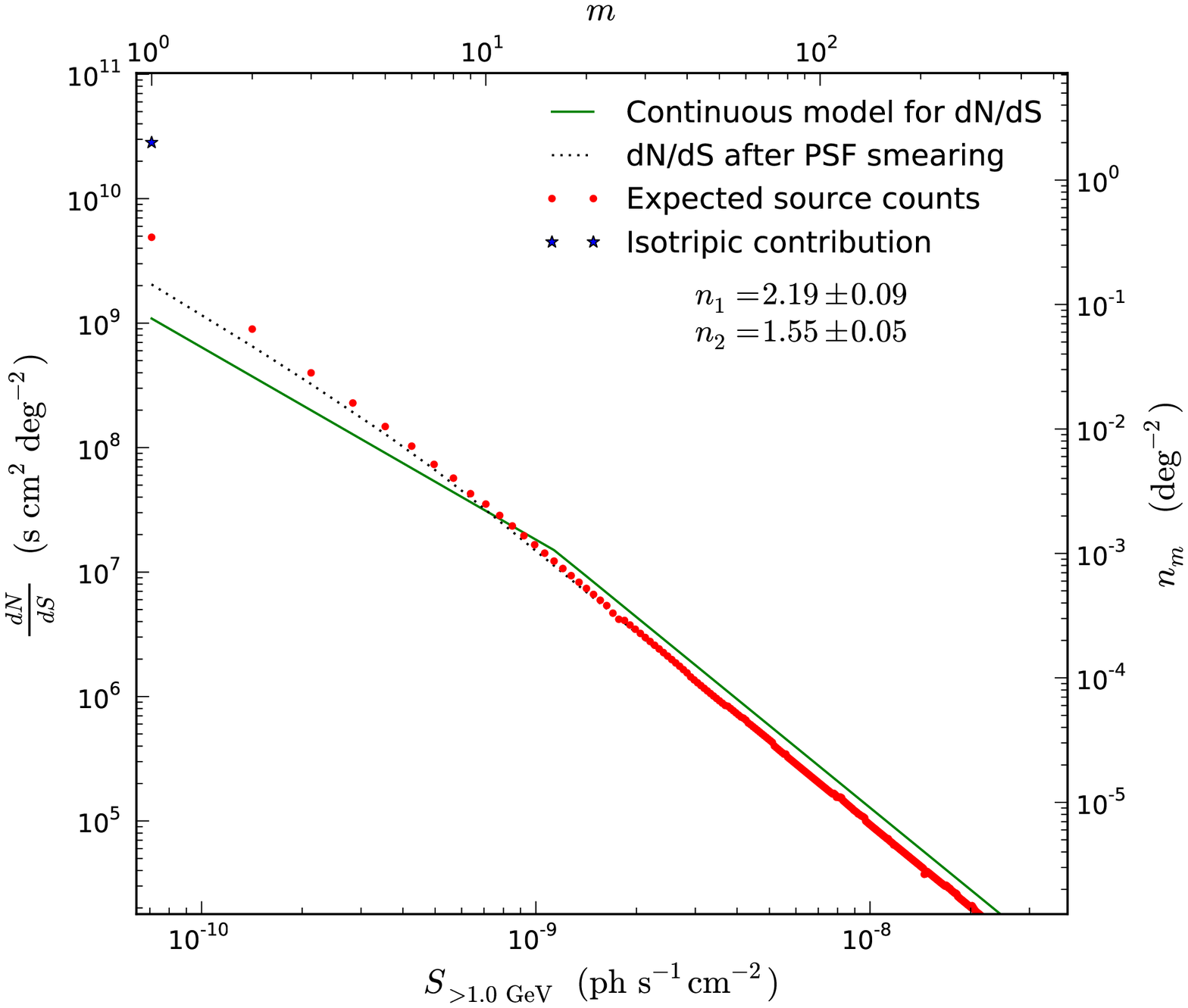, scale=\picsizetwo}
\end{center}
\vspace{-8mm}
\noindent
\caption{\small
Same as in Figure \ref{fig:datafit} but with the pixel size of about $1^\circ$
corresponding to HEALPix parameter nside = 64.
}
\label{fig:nside64}
\vspace{1mm}
\end{figure}

\begin{figure}[t] %[htbp] here, top, bottom, page
\begin{center}
\epsfig{figure = 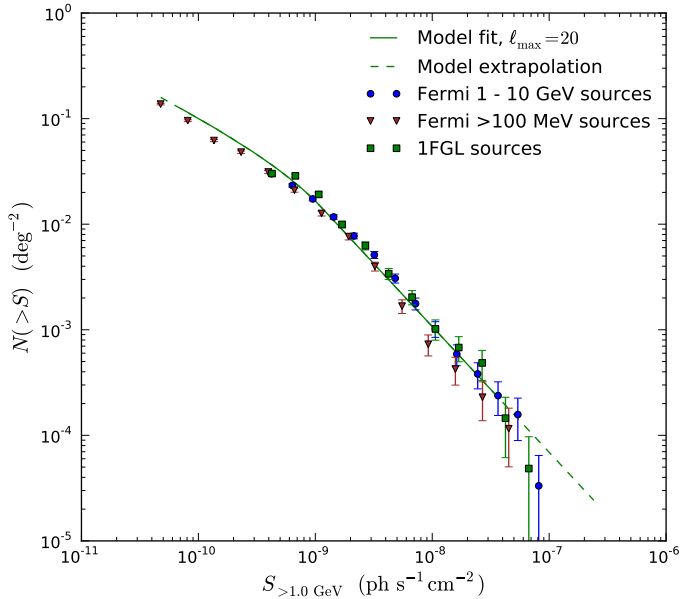, scale=\picsize} 
\end{center}
\vspace{-8mm}
\noindent
\caption{\small
Comparison with \Fermi source counts \citep{2010ApJS..188..405A, 2010ApJ...720..435A} 
in the case of analysis with HEALPix parameter nside = 64 (Figure \ref{fig:nside64}).
}
\label{fig:nside64_Fermi}
\vspace{1mm}
\end{figure}

The results for nside = 64 are presented in Figure \ref{fig:nside64}.
The total number of pixels inside the window below and above $30^\circ$ in latitude
is about 25,000,
i.e., the statistics of pixel counts is relatively good,
but the effect of PSF becomes more significant
than in the case of nside = 32.
For PSF = $0.4^\circ$, less than about 60\% of the flux from a source
can be inside one pixel.
In particular, one can note that the number of sources with small fluxes
is about two times larger than in the input model.
Nonetheless, 
in the case of nside = 64 the best-fit model is similar to nside = 32 case.
The position of the break,
$S_{\rm break} = 1.1 \times 10^{-9}{\rm ph\: s^{-1} cm^{-2}}$,
is somewhat lower
than the values in Table \ref{tab:compare}
but the overall source count distribution (Figure \ref{fig:nside64_Fermi})
is consistent with the source counts found by \Fermi collaboration 
\citep{2010ApJS..188..405A, 2010ApJ...720..435A}.

\begin{figure}[t] %[htbp] here, top, bottom, page
\begin{center}
\epsfig{figure = 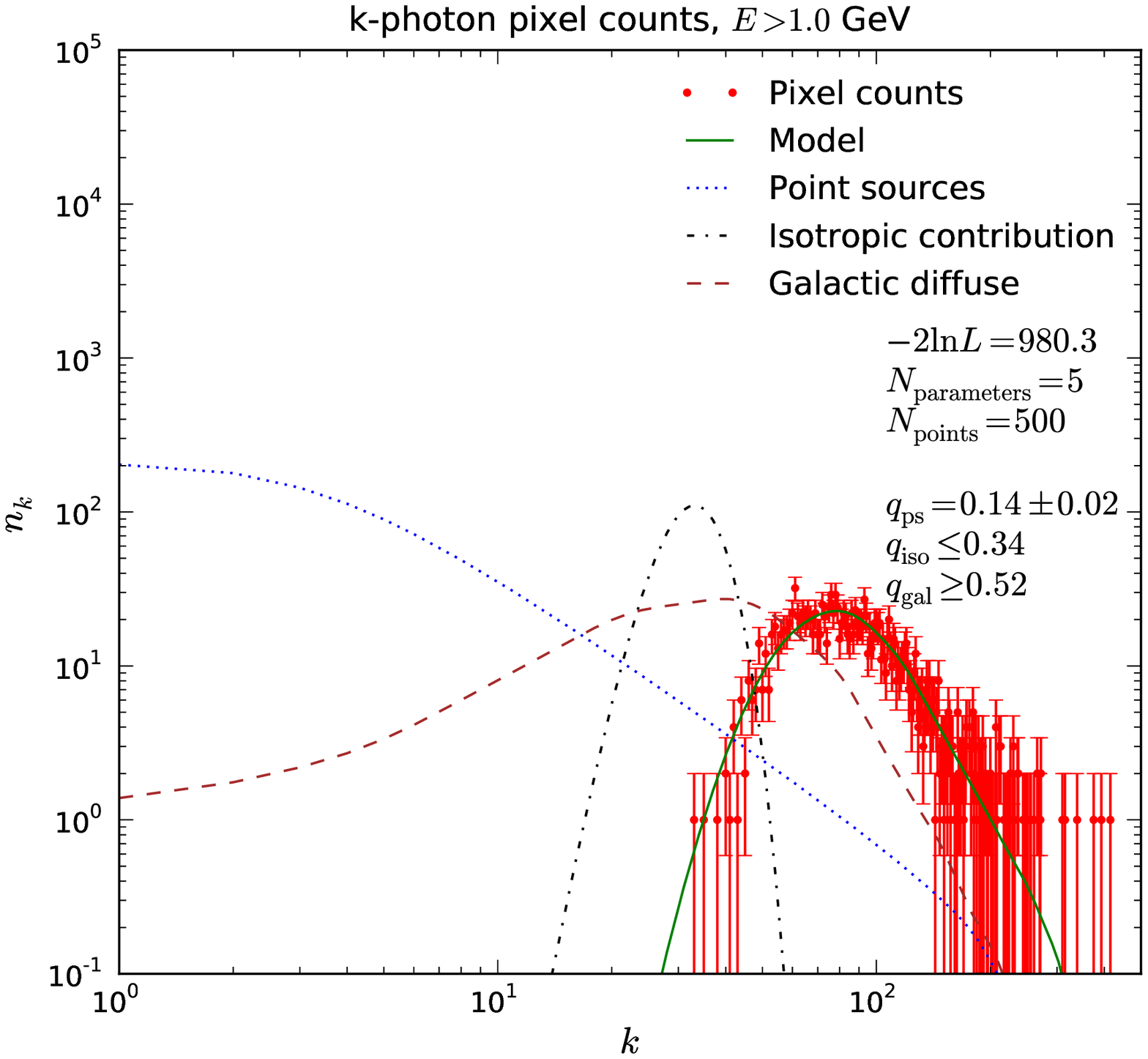, scale=\picsizetwo} 
\epsfig{figure = 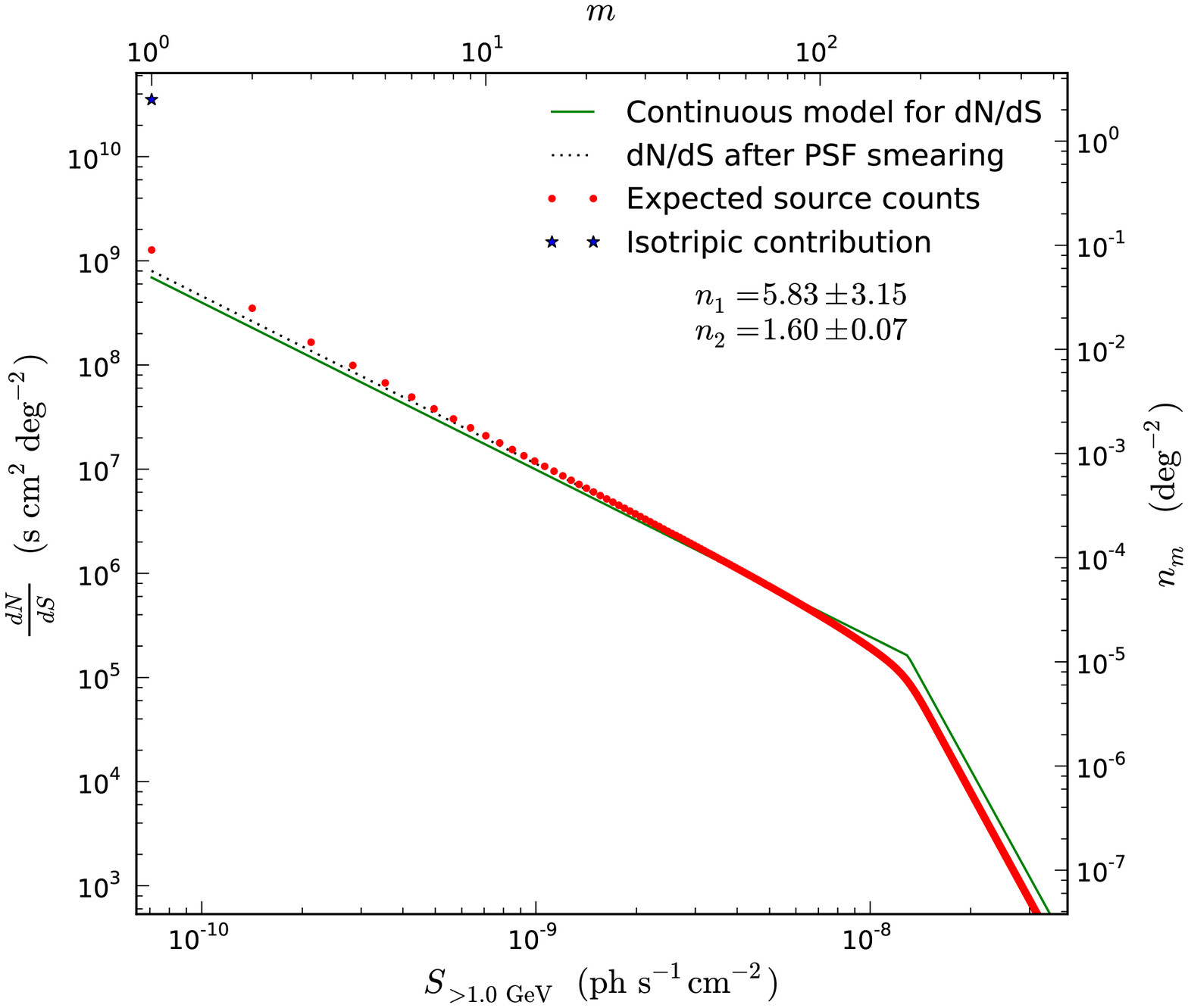, scale=\picsizetwo}
\end{center}
\vspace{-8mm}
\noindent
\caption{\small
Same as in Figure \ref{fig:datafit} but with the pixel size of about $4^\circ$
corresponding to HEALPix parameter nside = 16.
}
\label{fig:nside16}
\vspace{1mm}
\end{figure}

The results of fitting for nside = 16
are presented in Figure \ref{fig:nside16}.
In this case,  the PSF can be neglected.
However, the number of pixels inside the window is only about 1,500.
As one can see on the left side of Figure \ref{fig:nside16},
the error bars are rather large and the photon counts in most of the pixels are dominated
by the contribution from diffuse emission.
As a result, our method is not sensitive to the contribution from point sources
for nside = 16.
In particular, the MCMC is dominated by models with large $S_{\rm break}$.
The best fit value 
$S_{\rm break} = 1.3 \times 10^{-8}{\rm ph\: s^{-1} cm^{-2}}$
is an order of magnitude larger than the 
values in Table \ref{tab:compare}.
The reason is that models with large $S_{\rm break}$ have a comparable likelihood
to the models with small $S_{\rm break}$ but there are many more
ways to choose a large $S_{\rm break}$.

We conclude that our method is relatively robust to the change of pixel size:
the cases of nside = 32 and nside = 64 give very similar results,
unless the number of pixels is not sufficient to draw any conclusion 
about the statistics of sources, as in the case of nside = 16.

% end body
%:

\bibliography{PSpapers}         %or whatever your .bib file is

\end{document}